\documentclass[%
 aip,
 amsmath,amssymb,
 reprint,%
]{revtex4-1}

\usepackage{graphicx}
\usepackage{dcolumn}
\usepackage{bm}

\usepackage[utf8]{inputenc}
\usepackage[T1]{fontenc}
\usepackage{mathptmx}
\usepackage{etoolbox}
\usepackage{comment}
\usepackage{float}
\usepackage{adjustbox}
\usepackage{tabularx} 
\usepackage{xcolor}
\usepackage{tikz}
\usepackage{devanagari}

\makeatletter
\def\@email#1#2{%
 \endgroup
 \patchcmd{\titleblock@produce}
  {\frontmatter@RRAPformat}
  {\frontmatter@RRAPformat{\produce@RRAP{*#1\href{mailto:#2}{#2}}}\frontmatter@RRAPformat}
  {}{}
}%
\newcommand{\pfrac}[2]{\frac{\partial #1}{\partial #2}}

\makeatother

\makeatletter
\def\underbracex#1#2{\mathop{\vtop{\m@th\ialign{##\crcr
   $\hfil\displaystyle{#2}\hfil$\crcr
   \noalign{\kern3\p@\nointerlineskip}%
   #1\crcr\noalign{\kern3\p@}}}}\limits}

\def\upbracefilla{$\m@th \setbox\z@\hbox{$\braceld$}%
  \bracelu\leaders\vrule \@height\ht\z@ \@depth\z@\hfill 
\kern\p@\vrule \@width\p@\kern\p@\vrule \@width\p@\kern\p@\vrule \@width\p@
$}

\def\upbracefillb{$\m@th \setbox\z@\hbox{$\braceld$}%
\vrule \@width\p@\kern\p@\vrule \@width\p@\kern\p@\vrule \@width\p@\kern\p@
 \leaders\vrule \@height\ht\z@ \@depth\z@\hfill\bracerd
  \braceld\leaders\vrule \@height\ht\z@ \@depth\z@\hfill
\kern\p@\vrule \@width\p@\kern\p@\vrule \@width\p@\kern\p@\vrule \@width\p@
$}

\def\upbracefillc{$\m@th \setbox\z@\hbox{$\braceld$}%
\vrule \@width\p@\kern\p@\vrule \@width\p@\kern\p@\vrule \@width\p@\kern\p@
\leaders\vrule \@height\ht\z@ \@depth\z@\hfill
\kern\p@\vrule \@width\p@\kern\p@\vrule \@width\p@\kern\p@\vrule \@width\p@
$}

\def\upbracefilld{$\m@th \setbox\z@\hbox{$\braceld$}%
\vrule \@width\p@\kern\p@\vrule \@width\p@\kern\p@\vrule \@width\p@\kern\p@
 \leaders\vrule \@height\ht\z@ \@depth\z@\hfill\braceru$}

\def\upbracefillbd{$\m@th \setbox\z@\hbox{$\braceld$}%
\vrule \@width\p@\kern\p@\vrule \@width\p@\kern\p@\vrule \@width\p@\kern\p@
\bracerd\braceld
 \leaders\vrule \@height\ht\z@ \@depth\z@\hfill\braceru$}

\makeatother

\newif\ifmarkedup
\markedupfalse
\ifmarkedup
\newcommand{\Revision}[2]{\replaced{#2}{#1}}
\else
\newcommand{\Revision}[2]{{\textcolor{black}{#2}}}
\fi
\allowdisplaybreaks

\newcommand{\RevisionTwo}[1]{{\textcolor{black}{#1}}}

\begin{document}

\preprint{AIP/123-QED}

\title{Role of phase distortion in nonlinear saturation of the unstable acoustic modes in hypersonic parallel flow boundary layer}
\author{Altaf Ahmed}
\author{Joaquim P. Jossy}%

\author{Prateek Gupta}
\email{prgupta@iitd.ac.in}
\affiliation{%
Department of Applied Mechanics, Indian Institute of Technology Delhi, New Delhi 110016, India
}%

\date{\today}

\begin{abstract}
We analyze the role of the relative phasing in the nonlinear saturation of the unstable Mack modes in a hypersonic parallel flow boundary layer in two dimensions (2D). As the linearly unstable Mack modes extract energy from the mean flow, the perturbation energy cascades into higher harmonics as well as the mean flow. The higher harmonics are generated with $<0.5\%$ of total perturbation energy at steady state, indicating a very small role of higher harmonics in 2D. Additionally, the higher harmonics propagate with the same phase speed as the unstable mode, indicating wave steepening and a coherent energy cascade. The mean flow gets decelerated and heated due to the continuous extraction of the perturbation energy into traveling modes and the viscous dissipation of these modes. Unlike unstable modes in classical hydrodynamics, we show that the distortion in relative phasing between the streamwise velocity and wall-normal velocity due to nonlinear distortion of the mean flow is dominant. Using asymptotic reconstruction of the unstable eigenmodes, we compute the perturbation energy budgets in the linear and nonlinear regimes. Through energy budgets, we show that the viscous effects in the wall layer and the viscous effects in the critical layer sufficiently capture the distortion in phase due to the mean-flow distortion. 
We then combine this in a numerical model for calculating the steady-state perturbation energy and mean-flow distortion through the nonlinear saturation of unstable Mack modes in a hypersonic parallel flow boundary layer in 2D. Throughout, we compare the results of approximate theoretical analysis with 2D direct numerical simulations (DNS). 

\end{abstract}

\maketitle


\section{\label{sec:level1}Introduction}
  One of the primary mechanisms of laminar to turbulent transition in shear flows is the growth of primary instability followed by secondary breakdown~\cite{herbert1988secondary}. In incompressible and low to moderate Mach number compressible flat plate boundary layers, the most unstable primary instability modes are called the Tollmien-Schlichting modes (TS waves)~\cite{drazin2004hydrodynamic}. In high Mach number flat plate boundary layers ($M_\infty>4$, hypersonic), the most unstable primary instability modes have an acoustic signature, typically called the Mack modes~\cite{mack1969boundary, mack1975linear, mack1976numerical,mack1984boundary}. As the primary instability modes extract energy from the mean flow, the amplitudes become finite, resulting in significant nonlinear effects. In this work, we analyze and model the nonlinear saturation of Mack's acoustic instability modes in two dimensions. 

  The primary instability in boundary layers can be studied as a convective or an absolute instability~\cite{drazin2004hydrodynamic, schmid2002stability}. As an absolute instability or temporal instability, the temporal growth of modes at all spatial points is obtained as growth rates. However, as convective or spatial instability, the spatial growth rates of the modes are obtained as the modes are convected downstream. Assuming parallel flow, the temporal and spatial growth rates can be correlated~\cite{gaster1962note}. To account for the spatial growth, parabolic stability equations (linear and nonlinear~\cite{chang1993linear, bertolotti1991analysis, govindarajan1995stability}) are solved, assuming a WKB expansion in the streamwise wavenumber~\cite{saric1975nonparallel}. In both spatial and temporal linear stability analysis, \RevisionTwo{the modes are shown to be both acoustic as well as vortical}~\cite{sharma2023effect}. The acoustic waves mostly refract below the relative sonic line where the relative speed of the instability waves (with respect to base flow) is equal to the local speed of sound. Above this line, the waves become evanescent, decaying exponentially away from the wall. Before decaying completely, the instability modes exhibit strong thermoviscous effects~\cite{lees1946investigation, reshotko1960stability} near the critical line, where the relative speed of the instability waves (with respect to base flow) is zero. The production of the perturbation energy due to instability is a product of the relative phasing between the streamwise and wall-normal perturbation velocities ~\cite{reshotko1976boundary, hanifi1996transient}. Although the acoustic waves have high amplitude below the relative sonic line, the perturbation energy production takes place throughout the boundary layer.
  
  As pointed out by~\citeauthor{herbert1988secondary}\cite{herbert1988secondary}, the nonlinear mechanisms of boundary layer growth saturation depend strongly on the mean flow distortion. While high-fidelity 3D DNS studies of transitioning hypersonic flat plate boundary layers have been done~\cite{krishnan2006effect, sivasubramanian2010numerical, franko2013breakdown, unnikrishnan2020linear}, theoretical analysis and mechanism of nonlinear saturation of Mack modes in flat plate boundary layers has received relatively less attention, particularly the role of distortion in production due to mean flow distortion. In the context of Mack modes, multiple works in shear layers~\cite{goldstein1989nonlinear} and boundary layers~\cite{leib1995nonlinear, wu2019nonlinear} have been considered to account for the critical layer nonlinearity. The critical layer is a layer of thickness $\sim \mathrm{Re}^{-1/3}$, where $\mathrm{Re}$ denotes a suitably defined Reynolds number used in the dimensionless equations, around the critical line. Both the nonlinear and thermoviscous effects have been shown to compete in the critical layers~\cite{haberman1972critical}. Using the nonlinearity in the critical layer, the amplitude equation for the instability modes in the nonlinear regime can be derived. However, in this work, we show that linear thermoviscous effects in the critical layer are sufficient to capture the reduction in perturbation energy production due to the mean flow distortion. 
  The role of nonlinear distortion of the mean flow in saturation of linearly unstable modes in shear flows was first pointed out by~\citeauthor{stuart1958non}\cite{stuart1958non}, followed by a series of works on nonlinear saturation in internal shear flows~\cite{stuart1960non, watson1960non, stewartson1971non}. However, in all these works, the primary saturation mechanism is assumed to be the mean flow distortion, leading to a less efficient transfer of energy from the mean flow to the perturbations, assuming the shape and relative phasing of perturbations remain constant as the perturbations evolve. ~\citeauthor{stuart1958non}\cite{stuart1958non} derived a Landau equation for the amplitude, 
  \begin{equation*}
  \frac{dy}{d\tau} = \beta_1 y  - \beta_2 y  - \beta_3 y^2,
  \end{equation*}
  where $y$ is the amplitude square, which is a measure of the perturbation energy, and $\tau$ is the slow time scale (see Eq. 2.15 in ~\citeauthor{stuart1958non}\cite{stuart1958non}). The $\beta_3$ term in the above equation arises due to the interaction of distorted mean flow with the relative phase term between streamwise and wall-normal velocity modes, \emph{assuming the modes have the identical shape as the modes on undistorted laminar flow}. In doing so, the interactions of the distorted modes with the undistorted laminar flow are ignored. In this work, we show that this results in an overestimation of the mean flow distortion. Through 2D DNS, we show that the distortion of modes due to the mean flow distortion is the primary mechanism for the nonlinear saturation of an unstable Mack mode in a 2D hypersonic boundary layer parallel flow. In spatially developing hypersonic boundary layer experiments, mean-flow interaction has been observed to precede higher harmonic generation~\cite{chokani2005nonlinear}. \Revision{}{\citeauthor{zhu2020nonlinear}\cite{zhu2020nonlinear} study the nonlinear evolution of instability modes over the flared cone with permeable wall and compare it with hard wall case using LST, nonlinear PSE, and Floquet analysis. Porous walls have been shown to affect (delay) the transition of hypersonic boundary layers~\cite{fedorov2003stabilization, sousa2024dynamic} due to the interaction of the unstable acoustic modes with finite impedance boundaries~\cite{patel2018impedance}. \citeauthor{zhu2022instability} \cite{zhu2022instability} perform experiments and DNS over a flared cone with a wavy wall. Both studies~\cite{zhu2020nonlinear, zhu2022instability} investigate the effect of mean flow modified by wall boundary conditions on instability evolution.} While the impact of general externally imposed distortions on the Mack modes has been studied using adjoint-based sensitivity analysis~\cite{park2019sensitivity}, the impact of distortions on the Mack modes that occur due to the nonlinear propagation of the amplified Mack modes has not been considered. In this work, we analyze the modification in the perturbation energy production as the mean flow gets distorted due to the nonlinear propagation of the amplified mode.

In Sec.~\ref{sec: GovEqnsNumModel}, we discuss the dimensionless governing equations used in the 2D DNS and the linear stability analysis. We conduct 2D periodic flow simulations by adding body force and heating to maintain parallel flow with a flow profile identical to the self-similar solution of a spatially developing boundary layer. This setup has been used previously, often called temporal DNS~\cite{erlebacher1990numerical}. In Sec.~\ref{sec: results}, we discuss the modeling in linear and nonlinear regimes. We derive the linear perturbation energy budgets and show the variation of energy production and dissipation along the boundary layer. Additionally, we highlight the significance of viscous effects near the wall and the critical layer in capturing energy production in the linear regime. We then discuss the nonlinear saturation regime. We show that the nonlinear propagation of the Mack mode generates higher harmonics that are coherent and propagate at the same phase speed as the unstable mode. The nonlinear propagation also results in mean-flow distortion. Ultimately, we show that keeping the mode shape fixed and only accounting for mean flow distortion in energy production overestimates the distortion significantly while accounting for distortion in modes results in a correct prediction of the mean flow distortion, before concluding in Sec.~\ref{sec: conclusions}.

\section{Governing Equations and Numerical modeling}
\label{sec: GovEqnsNumModel}
The dimensional compressible Navier–Stokes equations for an ideal calorically perfect gas are,

\begin{subequations}\label{eq: Navier-Stokes dimensional}
\begin{align}\label{eq: continuity dimensional}
    &\frac{\partial \rho^{\star}}{\partial t^{\star}} + \nabla \cdot\left(\rho^{\star} \boldsymbol{u}^{\star}\right) =0,
\end{align}
\begin{align}\label{eq: momentum dimensional}
     &\rho^{\star}\left(\frac{\partial \boldsymbol{u}^{\star}}{\partial t^{\star}} + \left(\boldsymbol{u}^{\star} \cdot \nabla\right)\boldsymbol{u}^{\star}\right) =-\nabla p^{\star} + \nabla \cdot \left(2\mu^{\star} \boldsymbol{S}^{\star} \right.\nonumber\\
     &\left.+\left(\kappa^{\star}-\frac{2}{3}\mu^{\star}\right) \nabla\cdot\boldsymbol{u}^{\star}\boldsymbol{I}\right) + \boldsymbol{F}^{\star},
\end{align}
\begin{align}\label{eq: internal dimensional}
    &\rho^{\star}\left(\frac{\partial e^{\star}}{\partial t^{\star}} + \left(\boldsymbol{u}^{\star} \cdot \nabla\right)e^{\star}\right)=-p^{\star}\nabla\cdot \boldsymbol{u}^{\star} +2\mu^{\star} \boldsymbol{S}^{\star}:\boldsymbol{S}^{\star} +\nonumber\\&\left(\kappa^{\star}-\frac{2}{3}\mu^{\star}\right) (\nabla\cdot\boldsymbol{u}^{\star})^2 + \nabla \cdot( k^{\star}\nabla T^{\star}) + Q^{\star}_e, 
\end{align}
\begin{align}\label{eq: eos dimensional}
    p^{\star} = \rho^{\star} R^{\star} T^{\star},
\end{align}
\end{subequations}
where $\boldsymbol{S}^{\star}$ is the strain rate tensor defined as
\begin{align*}
    \boldsymbol{S}^{\star} = \frac{1}{2}\left(\nabla \boldsymbol{u}^{\star} +\nabla \boldsymbol{u}^{\star T}\right).
\end{align*}
Quantities $\mu^{\star}$, $\kappa^{\star}$, and $k^{\star}$ are dynamic viscosity, bulk viscosity, and thermal conductivity coefficients, respectively, and $\boldsymbol{F}^{\star}$ and $Q^{\star}_e$ denote the source terms. We use the superscript $()^{\star}$ to denote the dimensional quantities. As discussed later, we assume parallel flow in the analysis as well as the DNS. To maintain parallel flow with velocity and temperature profiles matching the steady state boundary layer profile at a given $x$ location, forcing in $x$ direction ${F}^{\star}_x$ and external heating $Q^{\star}_e$ are used. We assume bulk viscosity $\kappa^{\star}$ to be zero. To non-dimensionalize Eqs.~\eqref{eq: Navier-Stokes dimensional}, we consider the following dimensionless variables, 
\begin{align}
    &x =\frac{x^{\star}}{\delta^{\star}}, \hspace{5mm} y =\frac{y^{\star}}{\delta^{\star}}, \hspace{5mm} t= t^{\star}\frac{U^{\star}_{\infty}}{\delta^{\star}},\hspace{5mm}\rho = \frac{\rho^{\star}}{\rho^{\star}_{\infty}}, \hspace{5mm} \boldsymbol{u} = \frac{\boldsymbol{u}^{\star}}{U^{\star}_{\infty}},\nonumber \\  
    & T = \frac{T^{\star}}{T^{\star}_{\infty}} , \hspace{5mm} p = \frac{p^{\star}}{\rho^{\star}_{\infty}R^{\star}T^{\star}_{\infty}},\hspace{5mm} \mu = \frac{\mu^{\star}}{\mu^{\star}_{\infty}}, \hspace{5mm} k = \frac{k^{\star}}{k^{\star}_{\infty}},
\end{align}
where the subscript $()_{\infty}$ denotes the free stream quantities and $\delta^{\star}= \int \limits_{0}^{\infty}\left(1 - \frac{\rho^{\star}U^{\star}}{\rho^{\star}_{\infty}U^{\star}_{\infty}}\right) dy^{\star}$ is the displacement thickness. The resulting dimensionless governing equations are, 
\begin{subequations}\label{eq: Navier-Stokes Non-dimensional}
\begin{align}\label{eq: continuity non-dimensional}
    &\frac{\partial \rho}{\partial t} + \nabla \cdot\left(\rho \boldsymbol{u}\right) =0,
\end{align}
\begin{align}\label{eq: momentum non-dimensional}
     &\rho\left(\frac{\partial \boldsymbol{u}}{\partial t} + \left(\boldsymbol{u} \cdot \nabla\right)\boldsymbol{u}\right) = -\frac{1}{\gamma M^2_{\infty}}\nabla p + \frac{1}{\mathrm{Re}_{\delta^{\star}}}\nabla \cdot \left(2\mu \boldsymbol{S} \right.\nonumber\\&\left.-\frac{2}{3}\mu \left(\nabla\cdot\boldsymbol{u}\right)\boldsymbol{I}\right)+\boldsymbol{F},
 \end{align}
\begin{align}\label{eq: Temperature non-dimensional}
    &\rho\left(\frac{\partial T}{\partial t} + \left(\boldsymbol{u} \cdot \nabla\right)T\right) + (\gamma-1)p \nabla \cdot\boldsymbol{u} =\frac{2\gamma(\gamma-1) M^2_{\infty}}{\mathrm{Re}_{\delta^{\star}}}\mu\Big(\boldsymbol{S}:\boldsymbol{S} \nonumber\\
    & -\frac{1}{3}\left(\nabla\cdot\boldsymbol{u}\right)^2 \Big)+ \frac{\gamma}{\mathrm{Re}_{\delta^{\star}} \mathrm{Pr}}\nabla \cdot( k\nabla T) + Q_e, 
\end{align}
\begin{align}\label{eq: eos non-dimensional}
    p = \rho T,
\end{align}
\label{eq: non_dim_full_gov}
\end{subequations}
where $\gamma$,  $M_{\infty}$, $\mathrm{Re}_{\delta^{\star}}$, and $\mathrm{Pr}$ denote the specific heat ratio, the freestream Mach number, the  Reynolds number,  and the Prandtl number, respectively, and are given by,
\begin{align*}
    M_{\infty} = \frac{U^{\star}_{\infty}}{\sqrt{\rho^{\star}_{\infty}R^{\star}T^{\star}_{\infty}}}, \hspace{5mm} \mathrm{Re}_{\delta^{\star}} = \frac{\rho^{\star}_{\infty}U^{\star}_{\infty}\delta^{\star}}{\mu^{\star}_{\infty}}, \hspace{5mm} \mathrm{Pr} = \frac{\mu^{\star}_{\infty}C^{\star}_P}{k^{\star}_{\infty}}.
\end{align*}
The freestream quantities $()_{\infty}$ are fixed and do not vary with streamwise direction, and we choose $\gamma=1.4$, $\mathrm{Pr}=0.7$ and $\mathrm{Re}_{\delta^{\star}}=10000$ for all of our cases. Below, we outline the linearization of Eqs.~\eqref{eq: non_dim_full_gov} and the details of the 2D DNS used in this work.
\subsection{Linear Stability Analysis}
We perform the linear stability analysis within the parallel flow assumption (base velocity is in the $x$ direction, varying only in $y$) for two-dimensional perturbations. Additionally, we assume the base state temperature to vary in the $y$ direction only. We also consider non-dimensional dynamic viscosity and thermal conductivity to vary according to Sutherlands's law of viscosity,
\begin{equation}
 \mu = k = T^{\frac{3}{2}}\left(\frac{S + 1}{S + T}\right),
\end{equation}
where $S=1.8$. We decompose the instantaneous variables and diffusive properties ($\mu$ and $k$) into base-state (steady) and two-dimensional perturbations for the linear stability analysis,
\begin{align}
 \phi = \phi_0(y) + \epsilon_a \phi'(x,y,t),
 \label{eq: amplitude_decomp}
\end{align}
where $\phi = \left(u, v, p, \rho, T, \mu, k\right)$, $u$ is the $x$ direction (streamwise) velocity and $v$ is the $y$ direction (wall-normal) velocity. We use the base state variables $u_0 = f(y)$ and $T_0 = g(y)$ to define the complete base state as, 
\begin{equation}
 v_0 = 0,~p_0=1,~\rho_0 = \frac{1}{g},~\mu_0=k_0=g^{3/2}\left(\frac{S + 1}{S + g}\right).
\end{equation}
Substituting the expansion in Eq.~\eqref{eq: amplitude_decomp} in Eqs.~\eqref{eq: Navier-Stokes Non-dimensional} and ignoring terms of order $\mathcal{O}\left(\epsilon^2_a\right)$ or higher, we obtain the linearized Navier-Stokes equations (LNSE), 
\begin{subequations}
\begin{align}
 &\pfrac{\rho'}{t} + f\pfrac{\rho'}{x} + \frac{1}{g}\left(\pfrac{u'}{x} + \pfrac{v'}{y}\right) - \frac{v'}{g^2}\frac{dg}{dy} = 0,\\
 &\pfrac{u'}{t} + f\pfrac{u'}{x} + v'\frac{df}{dy} + \frac{g}{\gamma M^2_\infty}\pfrac{p'}{x} = \frac{g}{\mathrm{Re}_{\delta^{\star}}}\Bigg[\pfrac{}{y}\left(\mu_0\pfrac{u'}{y} + \mu'\frac{df}{dy}\right)\nonumber\\& + \frac{4}{3}\pfrac{}{x}\left(\mu_0\pfrac{u'}{x}\right) + \pfrac{}{y}\left(\mu_0\pfrac{v'}{x}\right) - \frac{2}{3}\pfrac{}{x}\left(\mu_0\pfrac{v'}{y}\right)\Bigg],\\
 &\pfrac{v'}{t} + f\pfrac{v'}{x} + \frac{g}{\gamma M^2_\infty}\pfrac{p'}{y} = \frac{g}{\mathrm{Re}_{\delta^{\star}}}\Bigg[\frac{4}{3}\pfrac{}{y}\left(\mu_0\pfrac{v'}{y}\right)\nonumber\\
 &+\pfrac{}{x}\left(\mu_0\pfrac{v'}{x}\right) + \pfrac{}{x}\left(\mu_0\pfrac{u'}{y} + \mu'\frac{df}{dy}\right) - \frac{2}{3}\pfrac{}{y}\left(\mu_0\pfrac{u'}{x}\right)\Bigg],\\
 &\pfrac{T'}{t} + f\pfrac{T'}{x} + v'\frac{dg}{dy} + g(\gamma - 1)\left(\pfrac{u'}{x} + \pfrac{v'}{y}\right)=\nonumber\\& \frac{\gamma g}{Pr \mathrm{Re}_{\delta^{\star}}}\Bigg[\pfrac{}{x}\left(\mu_0\pfrac{T'}{x}\right) + \pfrac{}{y}\left(\mu_0\pfrac{T'}{y}+\mu'\frac{dg}{dy}\right)\Bigg] + \nonumber\\
 &\frac{\gamma(\gamma - 1)M^2_\infty}{\mathrm{Re}_{\delta^{\star}}}g\frac{df}{dy}\left(2\mu_0\left(\pfrac{u'}{y} + \pfrac{v'}{x}\right) + \mu'\frac{df}{dy}\right),
\end{align}
\label{eq: LSA_gov}
\end{subequations}
and the linearized equation of state,
\begin{equation}
 p' = g\rho' + \frac{1}{g}T'.
\end{equation}
The terms composed of only base-flow get canceled in Eqs.~\eqref{eq: LSA_gov} by the forcing terms in Eqs.~\eqref{eq: non_dim_full_gov}. For 2D linear stability analysis, we decompose the perturbations into modal form as,
\begin{equation}
\phi' = \hat{\phi}(y)e^{i\left(\alpha x - \omega t\right)} + \mathrm{c.c.}= \hat{\phi}(y)e^{i\alpha\left( x - c t\right)} + \mathrm{c.c.},
\end{equation}
where $\phi = \left(u, v, p, T\right)$. $\alpha$, $\omega$, and $c=\omega/\alpha$ are wavenumber, circular frequency, and phase speed, respectively. We note that $\rho'$ and $\mu'$ can be expressed in terms of $T'$ and $p'$ using the linearized equation of state and the linearized Sutherland's law. Substituting the above decomposition into the linearised perturbation equations results in a generalized eigenvalue problem of the form,
\begin{equation}
    \boldsymbol{A} \hat{\phi} = \omega \boldsymbol{B} \hat{\phi}.
    \label{eq: EGV}
\end{equation}
The coefficients of $\boldsymbol{A}$ and  $\boldsymbol{B}$ are given in  \citet{malik1990numerical} and are not reproduced here for brevity. We solve the eigenvalue problem in Eq.~\eqref{eq: EGV} using the pseudospectral collocation method with modified Chebyshev polynomials as the basis functions~\cite{boyd} which satisfy the boundary conditions,
\begin{equation}
    \hat{u} = \hat{v}   = \hat{T}  = 0 ,\hspace{10 mm}\hspace{2 mm} y= 0, \hspace{2 mm} y \to \infty.
    \label{eq: BC EV}
\end{equation}
The eigenvectors ($\hat{u}, \hat{v}, \hat{T}$) are reconstructed using the modified Chebyshev basis.
Additional details of the pseudospectral Chebyshev formulation are given by Hanifi et al.~\cite{hanifi1996transient}. The eigenvalue analysis yields a complex frequency $\omega = \omega_r + i\omega_i$ as a function of $\alpha$, the Reynolds number $\mathrm{Re}_{\delta^{\star}}$, and the base state profiles $f$ and $g$. The imaginary part $\omega_i$ signifies exponential temporal growth.

\subsection{Transient forced 2D DNS}

\begin{figure}
    \centering
    \includegraphics[width=\linewidth]{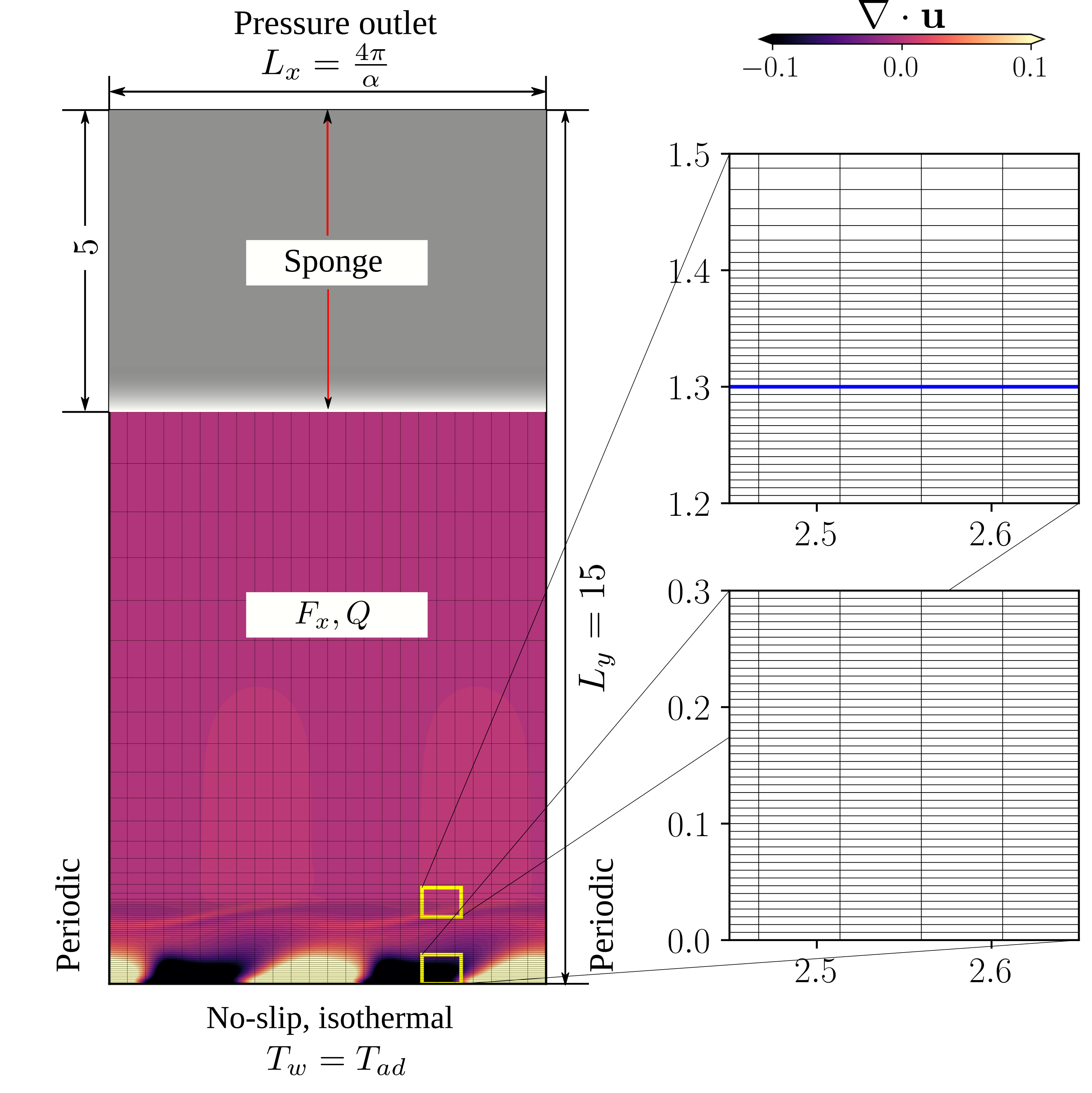}
    \put(-85,220){\small(i) near boundary layer}
    \put(-70,116){(ii) near wall}
    \caption{The rectangular computational domain with the inlet and outlet boundary set to be periodic; the bottom boundary is a no-slip isothermal wall with $T_w$ set to adiabatic wall temperature $T_{ad}$; at the top boundary, pressure outlet with a sponge layer between $y=10$ and $y=15$ is imposed. A zoomed-in view of the actual mesh near (i) wall and (ii) boundary layer edge is shown for reference.}
    \label{fig: computational_setup}
\end{figure}
\begin{table}[]
\caption{Flow and mesh parameter space of simulations used in this study. The \Revision{$\Delta y$}{$\Delta y^+$} denotes the $y$ direction spacing \Revision{}{in terms of friction units} of the uniform grid inside the boundary layer. The initial energy column corresponds to the initial perturbation energy prescribed.}
\begin{ruledtabular}

    \centering
        \begin{tabularx}{\linewidth}{p{10mm} p{10mm}p{10mm}p{15mm}p{15mm}p{25mm}} \\
        Case & $M_{\infty}$  & $\alpha$  & \Revision{}{$\Delta x^+$} &  \Revision{}{$\Delta y^+$} & Initial energy  \\[1mm]
        \hline \\[.5mm]
         $\mathrm{M1}$ & 4.5  & 2.25   & \Revision{0.0465}{0.9934} &  \Revision{0.0066}{0.1423} & $0$ \\ [1mm]
         $\mathrm{M1}^*$ & 4.5  & 2.25   & \Revision{0.0465}{0.9934} &  \Revision{0.0066}{0.1423} & $2.5 \times 10^{-7}$ \\ [1mm]
         $\mathrm{M1}^{**}$ & 4.5  & 2.25   &  \Revision{0.0465}{0.9934} &  \Revision{0.0066}{0.1423} &$2.5\times 10^{-5}$ \\ [1mm]
         M2 & 5.0  & 2.19   & \Revision{0.0478}{0.8903} &  \Revision{0.0066}{0.1241} & $0$  \\ [1mm]
         $\mathrm{M2}^*$  & 5.0  & 2.19   &\Revision{0.0478}{0.8903} &  \Revision{0.0066}{0.1241} &  $2.5\times 10^{-7}$\\ [1mm]
          $\mathrm{M2}^{**}$  & 5.0  & 2.19   & \Revision{0.0478}{0.8903} &  \Revision{0.0066}{0.1241} & $2.5\times 10^{-5}$ \\ [1mm]
         M3 & 5.5  & 2.14   & \Revision{0.0489}{0.8026} &  \Revision{0.0066}{0.1093} &  $0$  \\ [1mm]
          $\mathrm{M3}^*$  & 5.5  & 2.14   & \Revision{0.0489}{0.8026} &  \Revision{0.0066}{0.1093} &  $2.5\times 10^{-7}$  \\ [1mm]
         $\mathrm{M3}^{**}$  & 5.5  & 2.14   & \Revision{0.0489}{0.8026} &  \Revision{0.0066}{0.1093} & $2.5\times 10^{-5}$  \\ [1mm]
      
        \end{tabularx}
    \end{ruledtabular}

    \label{tab: parameters_table}
\end{table}

Due to the parallel flow assumption, we can study the nonlinear dynamics of the unstable mode in a relatively accessible computational setup. In 3D, the unstable 2D Mack modes result in fundamental and subharmonic resonances~\cite{qin2024excitation, zhao2023asymptotic} in the nonlinear regime. Since the focus of this work is on the nonlinear saturation of the 2D unstable Mack modes, we restrict our study to a two-dimensional setup. To this end, we perform temporal direct numerical simulations of a flat plate compressible boundary layer using an in-house spectral difference solver for compressible viscous Navier-Stokes equations~\cite{sahithi2024thermoviscous}. The solver has been validated with a spatially developing laminar hypersonic boundary layer (see supplementary information of~\citeauthor{sahithi2024thermoviscous}~\cite{sahithi2024thermoviscous}). Inviscid fluxes at element interfaces are calculated using the HLLC approximate Riemann solver~\cite{toro2013riemann, leveque2002finite}, and the viscous fluxes are calculated using extrapolation and averaging~\cite{jameson2014note}. We choose the streamwise length ($L_x$) such that $L_x=\frac{4\pi}{\alpha}= 2 \lambda$, where $\lambda$ is the wavelength of the fundamental instability mode. To keep the boundary layer free from external effects, we choose the transverse direction to be sufficiently larger\cite{erlebacher1990numerical} than the boundary layer displacement thickness $\delta^{\star}$. In all our simulations, we use $120 \times 301$  cells (excluding the sponge layer) for discretization in streamwise and transverse directions, respectively, where the variables in each cell are reconstructed using second-order Chebyshev polynomials in all directions. The cells are uniformly distributed along the streamwise direction, while we cluster $70\%$ of the cells in the transverse direction between $y=0$ and $y=1.4$. \Revision{}{The grid resolution in terms of friction units is given in Table \ref{tab: parameters_table}  and is defined as $\left(\Delta x^+, \Delta y^+\right) = \left(\Delta x, \Delta y \right)\sqrt{\frac{Re_{\delta^{\star}} \left.\frac{df}{dy}\right|_w \rho_w}{\mu_w}}$.} The details of the grid convergence study are given in appendix \ref{appendix: convergence}, and a representative sketch of the computational domain is shown in~Fig.~\ref{fig: computational_setup}. We use a no-slip isothermal boundary condition at the bottom wall (with temperature fixed at the adiabatic temperature) and periodic boundary conditions in the streamwise directions. To prevent any reflections in the domain, we use a sponge layer near the top boundary. To maintain parallel flow, the source term $F_x$, defined as, 
\begin{equation}
F_x = -\frac{1}{\mathrm{Re}_{\delta^{\star}}}\frac{d}{dy}\left(\mu_0\frac{df}{dy}\right),
\end{equation}
is added to the $x$ momentum equation. Since we solve for the conservative system of equations, the kinetic energy and the internal equations are added to obtain the total energy equation. Hence, the source term in the total energy equation becomes,
\begin{equation}
Q = Q_e + \gamma(\gamma - 1)M^2_{\infty}f F_x,
\end{equation}
where 
\begin{equation}
Q_e = -\frac{\gamma}{\mathrm{Re}_{\delta^{\star}}Pr}\frac{d }{d y}\left(\mu_0 \frac{d g}{d y}\right) -\frac{\mu_0\gamma\left(\gamma-1\right)M^2_{\infty}}{\mathrm{Re}_{\delta^{\star}}}\left(\frac{df}{dy}\right)^2.
\end{equation}

We conduct the DNS for three Mach numbers in the hypersonic range ($M_\infty = 4.5, 5.0, 5.5$) at $\mathrm{Re}_{\delta^{\star}}=10000$. For each Mach number, we consider three further cases, one in which we start with a quiescent steady flow. As the simulation proceeds, the most unstable mode (chosen based on $\alpha$) starts getting amplified. In the other two cases, we initialize the most unstable mode with some perturbation energy. Based on the steady-state perturbation energy, we choose these values as $2.5\times 10^{-7}$ (smaller than the saturated energy) and $2.5\times 10^{-5}$ (comparable to the saturated energy) (see Table~\ref{tab: parameters_table}). These cases help in confirming that the steady state limit cycle achieved is a stable limit cycle, within the range of initial perturbation energy values considered. Below, we discuss the dynamics in the linear regime and the nonlinear regime in detail. Everywhere, we use the case M1 (see Table~\ref{tab: parameters_table}) to showcase the results. However, the results and framework apply to all Mach number ranges in which the acoustic mode is unstable.

\begin{figure*}[!t]
    \centering
    \includegraphics[width=\linewidth]{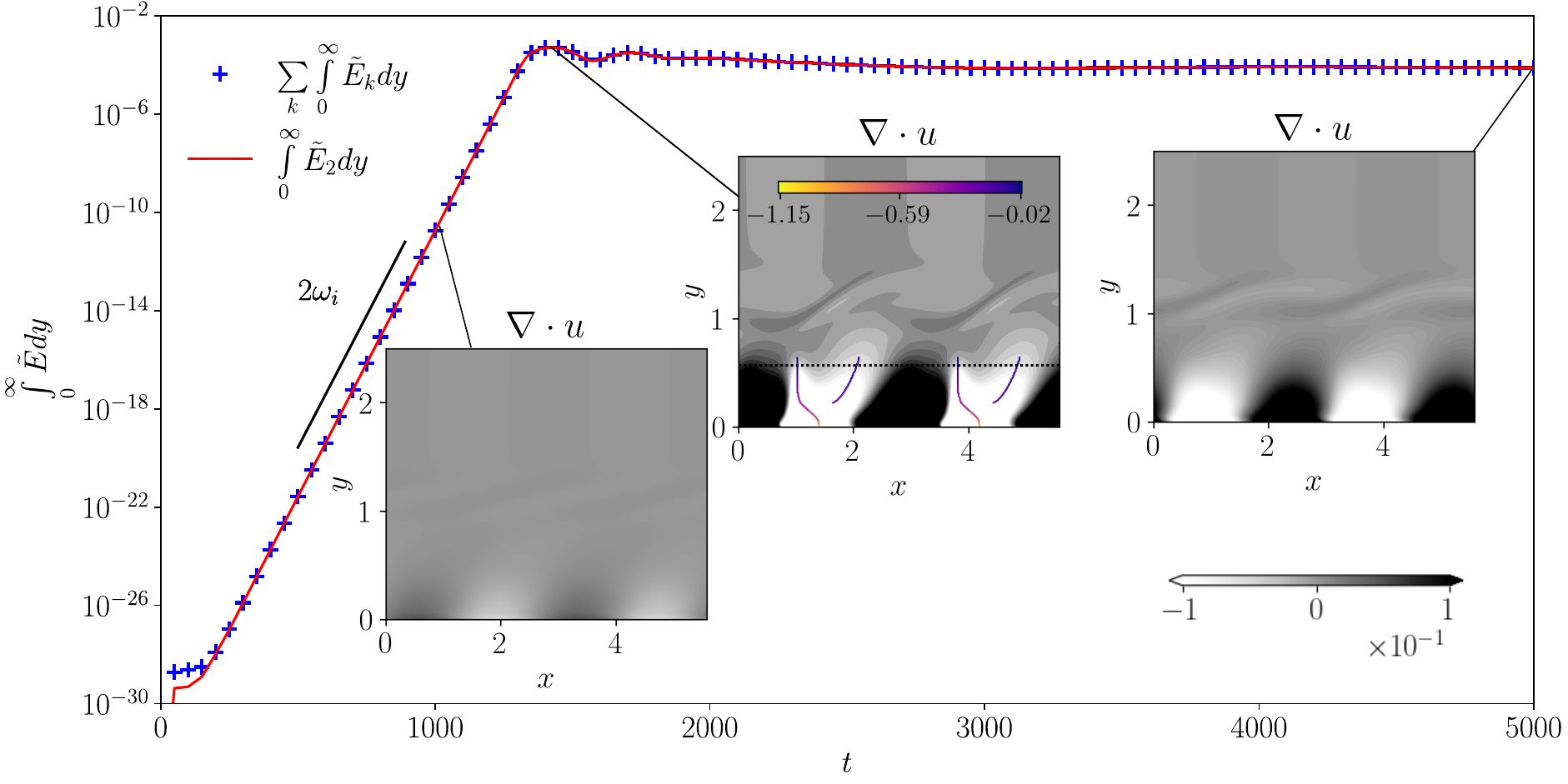}
    \put(-355,160){(i)~Linear}
    \put(-245,222){(ii)~Peak}
    \put(-105,222){(iii)~Steady}
    \caption{Time series of total perturbation energy in the second mode and combined for all modes along with the snapshots of velocity divergence at (i) linear growth stage, (ii) peak stage, and the (iii) nonlinear saturation stage for M1 case. Local minima of divergence at each $y$ are shown in (ii), colored by the corresponding divergence values. The highly negative values of divergence show steepened nonlinear waves of the traveling type mode.}
    \label{fig:perturbation-energy-semilog}
\end{figure*}

\section{Results and discussion}
\label{sec: results}

In this section, we discuss the results obtained from the 2D forced transient DNS  along with the linear stability analysis (LST). Using asymptotic approximations and solutions to the equations, we physically elucidate the nonlinear saturation observed in the DNS. Figure~\ref{fig:perturbation-energy-semilog} shows the growth of the perturbation energy in the second mode and all the modes combined for the M1 case (see Table~\ref{tab: parameters_table}) along with the visualization of velocity divergence in various regimes. The linear regime can be clearly identified by the exponential growth of the amplitude with the slope twice the growth rate ($2\omega_i$) obtained from LST. As the energy keeps increasing, two major nonlinear effects ensue, namely, the higher harmonic generation and the mean flow distortion. As discussed later, we show that the mean flow distortion results in the saturation of the perturbations. However, unlike incompressible flows~\cite{stuart1958non}, the saturation is due to the phase shift of the perturbations on the distorted mean flow.

\subsection{Linear regime}
\begin{figure}[!t]
    \centering
    \includegraphics[width=\linewidth]{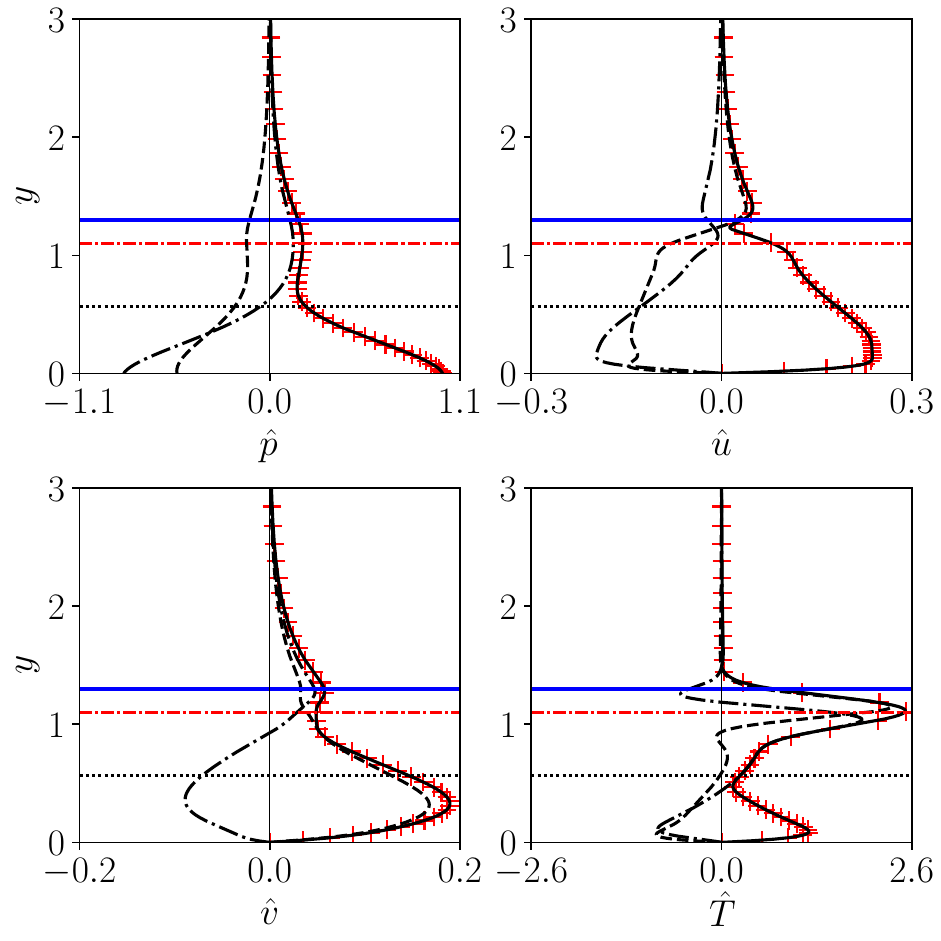}
    \put(-250,250){(a)}
    \put(-125,250){(b)}
    \put(-250,125){(c)}
    \put(-125,125){(d)}
    \caption{Eigenmodes obtained from LST (real part: $--$, imaginary part: $-~\cdot$, magnitude $-$ and DNS (red markers) for M1 case. The horizontal blue line is the boundary layer edge, the red chained line is the critical line, and the black dotted line is the relative sonic line. The complex frequency of the eigenmode from the LST is $2.049 + 0.0247i$ and from DNS is $2.049 + 0.0249i$.}
    \label{fig: eigenmodes_LST_DNS}
\end{figure}

In the linear regime, the second mode extracts energy from the parallel base flow and gets amplified exponentially, as has been studied extensively~\cite{fedorov2011transition}. The second mode contains both acoustic as well as vortical perturbations~\cite{sharma2023effect}. Very close to the wall, the acoustic perturbations are modified by the no-slip isothermal wall condition resulting in a thermoviscous Stokes' layer. As we move farther from the wall, the acoustic perturbations hit a so-called \emph{turning point} at $y$ where the speed of the flow w.r.t. perturbation waves ($f - c$) is the same as the local speed of sound $\sqrt{g}/M_\infty$. Most of the acoustic waves propagating away from the wall turn back towards the wall from the relative sonic line. Further away from the wall, the perturbations encounter a critical line at which the flow is stagnant w.r.t. perturbation waves $f=c$. Within a layer around the critical line, thermoviscous effects on perturbations are important~\cite{lees1946investigation}. As the perturbations keep getting amplified, both the nonlinear effects and the viscous effects compete within this critical layer~\cite{wu2019nonlinear}. Above the critical layer, all the modes decay exponentially as the flow quickly becomes a uniform parallel flow. 
To analyze the instability, perturbation energy budgets can be formulated by substituting the spatial modal decomposition $\left(\phi'= \Tilde{\phi}(y,t)e^{i\alpha x} + \mathrm{c.c}\right)$ in LNSE (Eqs.~\eqref{eq: LSA_gov}) to obtain,
\begin{subequations}
    \begin{align}
        & \frac{\partial \tilde{u}}{\partial t} + i\alpha f \tilde{u} + \tilde{v}\frac{df}{dy} + \frac{i\alpha g}{\gamma M^2_{\infty}}\tilde{p} =D_{\tilde{u}},\label{eq: LSA_x_mom_energy_budget}\\
         &\frac{\partial \tilde{v}}{\partial t} + i\alpha f \tilde{v} +  \frac{g}{\gamma M^2_{\infty}}\frac{\partial \tilde{p}}{\partial y} = D_{\tilde{v}}, \label{eq: LSA_y_mom_energy_budget}\\
         & \frac{\partial \tilde{p}}{\partial t} + i\alpha f \tilde{p} + \gamma \left(i\alpha \tilde{u} + \frac{\partial \tilde{v}}{\partial y}\right) = D_{\tilde{p}},\label{eq: LSA_pressure_energy_budget} 
    \end{align}
\end{subequations}where, 
\begin{subequations}
    \begin{align}
     &D_{\tilde{u}}= \frac{g}{\mathrm{Re}_{\delta^{\star}}}\left[ \frac{\partial}{\partial y}\left(\mu_0\frac{\partial \tilde{u}}{\partial y} + \frac{d\mu_0}{dg}\tilde{T}\frac{df}{dy}\right)- \frac{4}{3}\alpha^2\mu_0\tilde{u}+ i\alpha\frac{\partial \left(\mu_0\tilde{v}\right)}{\partial y} \right . \nonumber \\
     &\left .-\frac{2}{3}i\alpha\mu_0\frac{\partial \tilde{v}}{\partial y}\right],\\
    &D_{\tilde{v}}= \frac{g}{\mathrm{Re}_{\delta^{\star}}}\left[ \frac{4}{3}\frac{\partial}{\partial y}\left(\mu_0 \frac{\partial \tilde{v}}{\partial y}\right) - \alpha^2 \mu_0 \tilde{v} + i\alpha \left(\mu_0\frac{\partial\tilde{u}}{\partial y} + \frac{d\mu_0}{dT}\tilde{T}\frac{df}{dy}\right) \right . \nonumber \\
    & \left . - \frac{2}{3}i\alpha \frac{\partial}{\partial y} \left(\mu_0\tilde{u}\right)\right],\\
    &D_{\tilde{p}}  = \frac{\gamma}{Pr\mathrm{Re}_{\delta^{\star}}}\left[2\frac{dk_0}{dy}\frac{\partial \tilde{T}}{\partial y} +
      k_0\left( -\alpha^2 \tilde{T} + \frac{\partial^2 \tilde{T}}{\partial y^2}\right) + \frac{d^2 k_0}{dg^2}\tilde{T}\left(\frac{dg}{dy}\right)^2 \right . \nonumber \\
      & \left. + \frac{dk_0}{dg}\tilde{T}\frac{d^2g}{dy^2}\right] + \frac{\gamma(\gamma-1)M^2_{\infty}}{\mathrm{Re}_{\delta^{\star}}} \left[2\mu_0 \frac{df}{dy}\left(\frac{\partial \tilde{u}}{\partial y}+ i\alpha \tilde{v}\right) \right . \nonumber \\
      & \left.+ \frac{d\mu_0}{dg}\tilde{T}\left(\frac{df}{dy}\right)^2\right].
\end{align}
\end{subequations}
To define a correct perturbation energy norm, we multiply the Eqs.~\eqref{eq: LSA_x_mom_energy_budget}, \eqref{eq: LSA_y_mom_energy_budget}, and \eqref{eq: LSA_pressure_energy_budget} with $\frac{\tilde{u}^*}{g}$,$\frac{\tilde{v}^*}{g}$, and $\frac{\tilde{p}^*}{\left(\gamma M_{\infty}\right)^2}$, respectively, add them along with their complex conjugates~\cite{george2011chu, gupta2018spectral}. Here, $^*$ denotes the complex conjugate of the variable. The result yields the time evolution equation of the perturbation energy in the linear regime,
\begin{align}
    &\frac{\partial \tilde{E}}{\partial t}  + \frac{1}{\gamma M^2_{\infty}}\frac{\partial}{\partial
    y}\Re\left(\tilde{p}\tilde{v}^*\right) =  \underbrace{-\frac{1}{g}\frac{df}{dy}\Re\left(\tilde{u}\tilde{v}^*\right)}_{\text{$\mathbb{P}$}} + \underbrace{\frac{1}{g}\Re\left(D_{\tilde{u}} \tilde{u}^* \right)}_{\mathrm{I}} +  \nonumber\\
    &\underbrace{\frac{1}{g}\Re\left(D_{\tilde{v}} \tilde{v}^* \right)}_{\mathrm{II}} +
    \underbrace{\frac{1}{\left(\gamma M_{\infty}\right)^2}\Re\left(D_{\tilde{p}} \tilde{p}^* \right)}_{\mathrm{III}} = \mathbb{P} + \mathbb{D},
    \label{eq: pertEnergyLinear}
\end{align}
where $\tilde{E}$ is the perturbation energy, defined as,
\begin{align}
    \tilde{E} = \frac{1}{2g} \left(|\tilde{u}|^2 + |\tilde{v}|^2 \right) + \frac{|\tilde{p}|^2}{2\left(\gamma M_{\infty}\right)^2}.
\end{align}
\begin{figure}[!b]
    \centering
    \includegraphics[width=\linewidth]{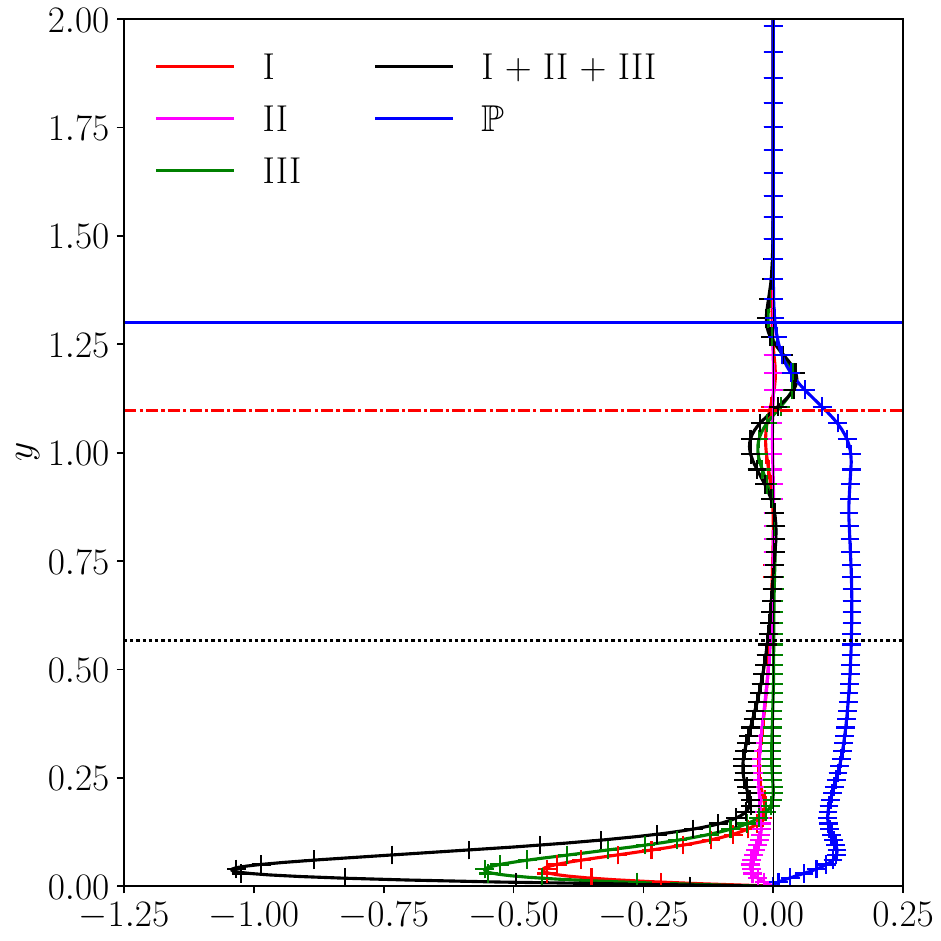}
    \caption{Variation of production and dissipation terms normalized by total energy $E^T = \int\displaylimits^\infty_0\tilde{E}dy$ in the perturbation energy equation \eqref{eq: pertEnergyLinear} along $y$ for the second mode for M1 case. The solid line `-' denotes LST, and the symbol `+' denotes DNS.}
    \label{fig: energy-balance-perturbation}
\end{figure}
In Eq.~\eqref{eq: pertEnergyLinear}, the underbraced terms are identified as the production term $\mathbb{P}$ and the dissipation terms $\mathrm{I},~\mathrm{II},~\mathrm{III}$ ($\mathbb{D}$=I+II+III). These terms are shown in Fig.~\ref{fig: energy-balance-perturbation} as obtained from the LST and the DNS calculations. The symbol $\Re()$ represents the real part of the quantity in the bracket. As evident from Fig.~\ref{fig: energy-balance-perturbation}, the production term dominates the dissipation terms above $y\gtrapprox 0.2$ within the boundary layer, indicating that the perturbation energy is generated within the boundary layer for all $y\gtrapprox 0.2$. Additionally, majority contribution within the dissipation is through the $\tilde{u}$ term ($\mathrm{I}$) and the $\tilde{p}$ term ($\mathrm{III}$). The production term $\mathbb{P}$ becomes slightly negative close to the boundary layer edge and the dissipation term $\mathbb{D}$ becomes slightly positive above the critical layer. While locally, these terms may cross 0, we label them production or dissipation based on the total values integrated in $y$.

\begin{figure}
    \centering
    \includegraphics[width=\linewidth]{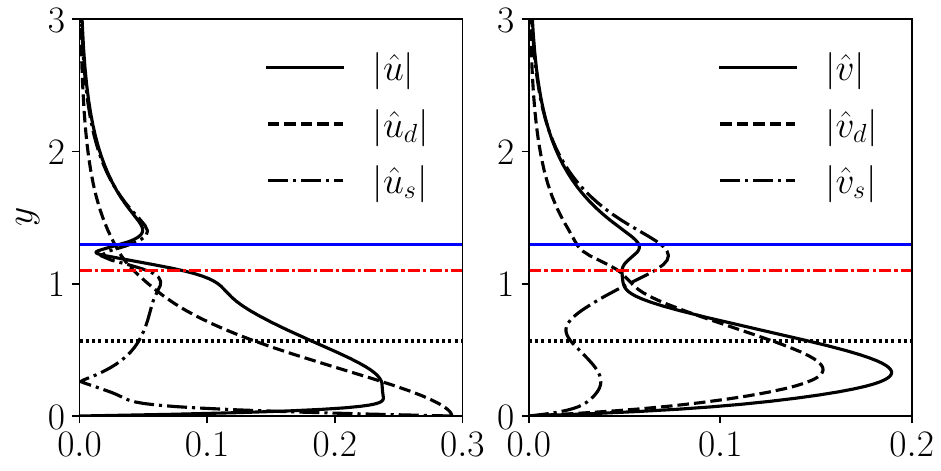}
    \put(-250,125){(a)}
    \put(-125,125){(b)}
    \caption{Dilatational and solenoidal components of velocity mode shape    of second mode for M1 case.}
    \label{fig: dilattion-solenoidal-velocity}
\end{figure}

The perturbation energy defined in Eq.~\eqref{eq: pertEnergyLinear} contains both the acoustic and the vortical perturbation energy. These components can be separated using the Helmholtz decomposition~\cite{sharma2023effect, jossy2023baroclinic}. Writing the velocity decomposition as,
\begin{equation}
    \boldsymbol{u} = \boldsymbol{u}_d + \boldsymbol{u}_s = \nabla \Phi + \nabla\times\boldsymbol{\psi},
\end{equation}
and imposing the divergence-free condition on $\boldsymbol{\psi}$ ($\nabla\cdot\boldsymbol{\psi} = 0$), we obtain,
\begin{equation}
    \nabla\cdot\boldsymbol{u} = \nabla\cdot\boldsymbol{u}_d = \nabla^2\Phi.
\label{eq: Helmholtz_2}
\end{equation}
Substituting the normal mode assumption in Eq.~\eqref{eq: Helmholtz_2}, we obtain,
\begin{equation}
    i\alpha\hat{u} + \frac{d\hat{v}}{dy} = \left(\frac{d^2}{dy^2} - \alpha^2\right)\hat{\Phi}.
    \label{eq: potential_poisson}
\end{equation}
At the hard-wall, the $y$ direction velocity vanishes imposing $d\hat{\Phi}/dy=0$ at $y=0$. Furthermore, far away from the boundary layer, all perturbations decay~\cite{sharma2023effect, hirasaki1970boundary}. Thus imposing $d\hat{\Phi}/dy=0$ for $y\to\infty$, we solve the ODE in Eq.~\eqref{eq: potential_poisson} using the modified Chebyshev basis, which satisfy the homogeneous Neumann conditions. Figure~\ref{fig: dilattion-solenoidal-velocity} shows the magnitude of the components of the dilatational and solenoidal velocity fields. Close to the wall, both the solenoidal and the dilatational fields peak. However, further from the wall, the solenoidal components decay rapidly. Close to the critical layer, the peak of the solenoidal component again highlights the importance of viscous effects within the critical layer in the linear regime.

Close to the wall, the acoustic perturbations result in a thermoviscous Stokes' layer. We refer to this regime as the inner layer. Away from the wall, the acoustic fluctuations are far away from being influenced by the no-slip isothermal conditions. In this regime, the perturbations are propagating with a continuous refraction from the varying thermal conditions and shear. We refer to this regime as the outer layer. In this section, we present an asymptotic derivation of mode shapes in these regimes. Using the LST calculations, DNS results, and these asymptotic mode shapes, we discuss the perturbation energy budgets.
\subsection{Asymptotic shape of eigenmodes}
\label{sec: asymptotic_shape_eigenmodes}

To derive the asymptotic governing equations in the inner and the outer layers, we scale the linear stability Eqs.~\eqref{eq: LSA_gov}. We use 
\begin{equation}
 \mathrm{Re}_{\delta^{\star}} = \frac{1}{\epsilon^2},
 \label{eq: Re_eps_scaling}
\end{equation}
for small parameter $\epsilon\ll 1$ and the perturbation expansions, 
\begin{equation}
 \hat{\phi} = \phi^{(0)} + \epsilon\phi^{(1)} + \mathcal{O}(\epsilon^2).
 \label{eq: asymptotic_expansion}
\end{equation}

\subsubsection{Wall layer solution}
\label{sec: wall_layer_solution}

Close to the wall, we stretch the wall-normal coordinate using the typical Stokes' layer $\delta_s$
\begin{equation}
\frac{\delta_s}{\delta^{\star}} \sim\frac{1}{\sqrt{\mathrm{Re}_{\delta^{\star}}}},
\end{equation}
which suggests the stretching of the form,
\begin{equation}
 y = \epsilon \xi,
  \label{eq: inner_layer_scaling}
\end{equation}
where $\xi\in[0,\infty)$ denotes the inner layer. For asymptotic matching, $\epsilon^{1-\lambda}\xi$ is kept constant as $\epsilon\to 0^+$ and $\xi\to \infty$ for $0<\lambda<1$~\cite{bender2013advanced, khamis2017acoustics}. Hence, the quantity $\epsilon \xi$ is of an intermediate order ($\mathcal{O}(\epsilon)<\epsilon \xi < \mathcal{O}(1)$). Converting the $y$ derivatives to $\xi$ derivatives in Eqs.~\eqref{eq: LSA_gov} and using the Taylor expansion of $f(y)$ and $g(y)$ close to the walls as,
\begin{subequations}
 \begin{align}
  &f(\epsilon \xi) = f'_w\epsilon \xi + \frac{f''_w}{2}\epsilon^2\xi^2\cdots,\\
  &g(\epsilon \xi) = g_w + \frac{g''_w}{2}\epsilon^2\xi^2\cdots.\label{eq: base_g_expansion}
 \end{align}
\end{subequations}
we obtain the inner layer modal continuity equation correct up to $\mathcal{O}(\epsilon)$ as
 \begin{align}
  \frac{1}{g_w}D v^{(0)} + \epsilon\left[i\Omega^{(0)}\rho^{(0)} + \frac{1}{g_w}\left(i\alpha u^{(0)} + D v^{(1)}\right)\right] = 0,
 \end{align}
where $\Omega^{(0)} = -\omega + \alpha f'_w\epsilon\xi$ and $D=\frac{d}{d\xi}$. In Eq.~\eqref{eq: base_g_expansion} we have used the condition that the flat plate is held at constant temperature equal to the adiabatic temperature. Using the hard-wall boundary condition at $\xi=0$, we obtain that $y$ direction velocity vanishes at the leading order, that is, $v^{(0)}=0$ within the inner layer. Thus, the continuity equation yields
\begin{equation}
 i\Omega^{(0)}\rho^{(0)} + \frac{1}{g_w}\left(i\alpha u^{(0)} + Dv^{(1)}\right) = 0
 \label{eq: inner_cont}
\end{equation}
Using Eqs.~\eqref{eq: Re_eps_scaling} and~\eqref{eq: inner_layer_scaling}, we obtain the following equations from $\mathcal{O}(1)$ terms in momentum and energy equations in Eqs.~\eqref{eq: EGV},
\begin{subequations}
 \begin{align}
  &i\Omega^{(0)}u^{(0)} + \frac{g_w}{\gamma M^2_\infty}i\alpha p^{(0)} = g_w\mu_w D^2u^{(0)},~\label{eq: inner_x_vel}\\
  &Dp^{(0)} = 0,\label{eq: inner_pressure}\\
  &i\Omega^{(0)} p^{(0)} + \gamma \left(i\alpha u^{(0)} + D v^{(1)}\right) = \frac{\gamma \mu_w}{Pr}D^2 T^{(0)}.\label{eq: inner_T}
 \end{align}
 \label{eq: inner_layer_open}
\end{subequations}
Using Eqs.~\eqref{eq: inner_cont}, \eqref{eq: inner_pressure}, and~\eqref{eq: inner_T} we obtain the inner layer density equation,
\begin{equation}
 i\Omega^{(0)}\left(\rho^{(0)} - \frac{p^{(0)}}{\gamma g_w}\right) = \frac{\mu_w g_w}{Pr}D^2\rho^{(0)},~\label{eq: inner_density}
\end{equation}
where we have used the leading order approximation of the linearized ideal gas equation
\begin{equation}
 p^{(0)} = \rho^{(0)}g_w + \frac{T^{(0)}}{g_w}.
\end{equation}
Equations~\eqref{eq: inner_x_vel} and~\eqref{eq: inner_density} can be solved using the Airy's functions as,
\begin{subequations}
 \begin{align}
&u^{(0)} = \frac{\left(i\eta\right)^{2/3}\alpha g_w p^{(0)}}{\gamma M^2_\infty\mathrm{Ai}'\left(-\left(i\eta \right)^{1/3}\omega\right)}{\int\displaylimits^{\Omega^{(0)}}_{-\omega}\mathrm{Ai}\left(\left(i\eta \right)^{1/3}\tau\right)d\tau},\\
&\rho^{(0)} = \frac{p^{(0)}}{\gamma g_w}\left(1 + (\gamma - 1)\frac{\mathrm{Ai}\left(\left(i\eta Pr\right)^{1/3}\Omega^{(0)}\right)}{\mathrm{Ai}\left(-\left(i\eta Pr\right)^{1/3}\omega\right)}\right),\\
&\eta = \frac{1}{g_w \mu_w \left(\epsilon \alpha f'_w\right)^2},
 \end{align}
\label{eq: inner_solution}
\end{subequations}
where the no-slip and isothermal boundary conditions have been used. Equations~\eqref{eq: inner_solution} contain corrections smaller than $\mathcal{O}(1)$ but larger than $\mathcal{O}(\epsilon)$. Strictly $\mathcal{O}(1)$ solution is obtained by replacing $\Omega^{(0)} = -\omega$ in Eqs.~\eqref{eq: inner_layer_open} as,

\begin{subequations}
\begin{align}
&u^{(0)} = -\frac{\alpha p^{(0)}g_w}{\omega \gamma M^2_\infty}\left(1 - e^{-\xi/\delta_s}\right),\\
&\rho^{(0)} = \frac{p^{(0)}}{\gamma g_w}\left(1 + (\gamma - 1)e^{-\xi\sqrt{Pr}/\delta_s}\right),\\
&\delta_s = \sqrt{-\frac{g_w\mu_w}{i\omega}}.
\end{align}
\label{eq: wall_layer_classic}
\end{subequations}
Equations~\eqref{eq: wall_layer_classic} yield the first-order solution in the wall layer. Below, we discuss the numerical calculation of the outer layer solution. The two solutions can be asymptotically matched using the $\xi\to\infty$ limit from Eqs.~\eqref{eq: wall_layer_classic}.

\subsubsection{Outer layer solution}
\label{sec: outer_layer_solution}
To formulate the outer solution, the $x$ and $y$ length scales may be taken to be similar. Using the asymptotic expansion in Eq.~\eqref{eq: asymptotic_expansion}, which yields the leading order outer layer equations as,
\begin{subequations}
 \begin{align}
  &i\Omega\rho^{(0)} + \frac{1}{g}\left(i\alpha u^{(0)} + \frac
  {d v^{(0)}}{dy}\right) - \frac{v^{(0)}}{g^2}\frac{dg}{dy} = 0,\\
  &i\Omega u^{(0)} + v^{(0)}\frac{df}{dy} + \frac{g}{\gamma M^2_{\infty}}i\alpha p^{(0)} = 0,\\
  &i\Omega v^{(0)} + \frac{g}{\gamma M^2_{\infty}}\frac{dp^{(0)}}{dy} = 0,\\
  &i\Omega p^{(0)} + \gamma\left(i\alpha u^{(0)} + \frac{dv^{(0)}}{dy}\right) = 0,
  \end{align}
 \label{eq: outer_layer_all}
\end{subequations}
combined with the equation of state,
\begin{equation}
 p^{(0)} = \rho^{(0)}g + \frac{T^{(0)}}{g}.
\label{eq: outer_EOS}
\end{equation}
Equations~\eqref{eq: outer_layer_all} govern inviscid propagation of acoustic waves and can be combined to obtain the Rayleigh pressure equation (also known as the Pridmore-Brown equation~\cite{reshotko1976boundary, rienstra2019solutions}),
\begin{equation}
         \frac{1}{\alpha^2}\frac{d^2 p^{(0)}}{dy^2}-Qp^{(0)} + \frac{Q'}{1 -Q}\frac{1}{\alpha^2}\frac{dp^{(0)}}{dy}=0,
\label{eq: pridmore-brown-outer}
\end{equation}
where $Q = 1 - \frac{M^2_\infty \left(f-c\right)^2}{g}$ and $Q' = dQ/dy$. We follow the procedure explained by Mack~\citep{mack1969boundary} by transforming Eq.~\eqref{eq: pridmore-brown-outer} to the nonlinear equation~\cite{reshotko1960stability, reshotko1976boundary} using the transformation $G = \frac{1}{\alpha^2 p^{(0)}}\frac{dp^{(0)}}{dy}$ which yields, 
\begin{equation}
\frac{dG}{dy} = Q + \frac{Q'}{1-Q}G - \alpha^2 G^2.
\label{eq: Riccati_outer}
\end{equation}
Far away from the boundary layer, the pressure perturbations decay as $p^{(0)}\sim e^{-\alpha\sqrt{Q_\infty}y}$ using which, we obtain the initial condition $G(y\to\infty)= -\frac{-\sqrt{Q_\infty}}{\alpha}$. Using this, we integrate Eq.~\eqref{eq: Riccati_outer} from freestream towards the wall. Near the critical line $Q=1$, the singularity is tackled by integrating in the complex plane by analytically extending $y$ and $G(y)$ and using the series solution of $G$ near the critical point. We refer the reader to appendix B of Reshotko's thesis~\cite{reshotko1960stability} and Mack's monograph~\cite{mack1969boundary} for details. Solving for $G(y)$ we obtain $p^{(0)}, u^{(0)}, v^{(0)} $, and $T^{(0)}$ using Eqs.~\eqref{eq: outer_layer_all}. 
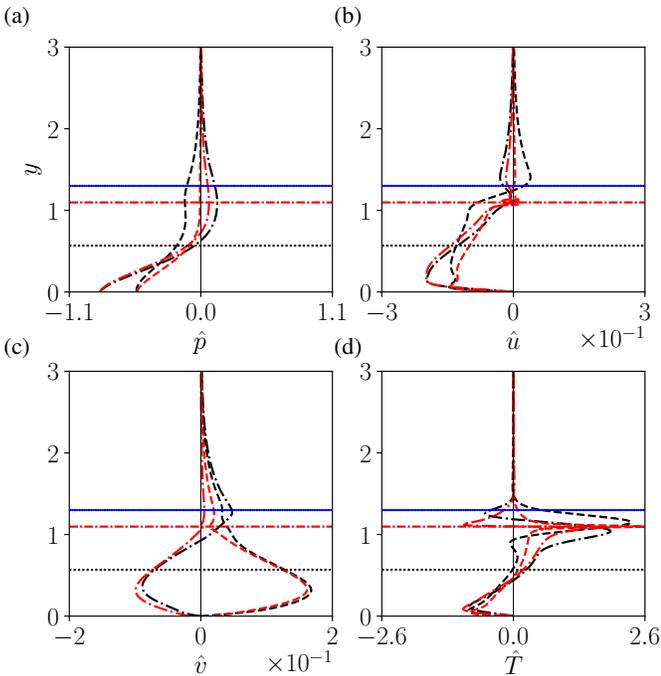
\begin{figure}[!b]
    \centering
    \includegraphics[width=\linewidth]{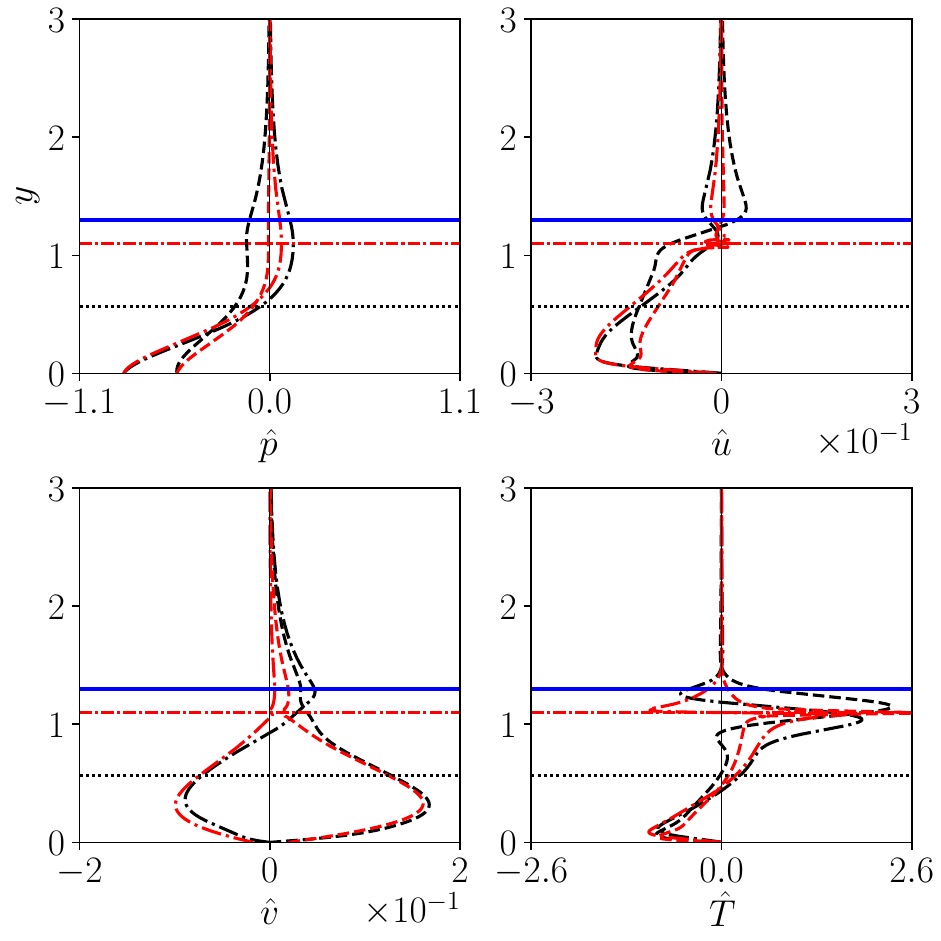}
    \put(-250,250){(a)}
    \put(-125,250){(b)}
    \put(-250,125){(c)}
    \put(-125,125){(d)}
    \caption{Comparison of the LST modes with the asymptotically matched modes using the inviscid solution and viscous wall-layer solution from Eqs.~\eqref{eq: wall_layer_classic} for M1 case. The black line denotes LST, and the red line denotes asymptotic solutions, with a dashed line showing the real part and a chained line showing the imaginary part.}
    \label{fig: composite_linear}
\end{figure}
Figure~\ref{fig: composite_linear} shows the comparison of the asymptotically matched linear solutions obtained from the inviscid solution and Eqs.~\eqref{eq: wall_layer_classic} with the LST modes. Given that we use only an inviscid solution in the critical layer, the modes exhibit oscillations in the critical layer. The solution is very well-behaved close to the turning point (at the relative sonic line) where $Q=0$. In the classic inviscid stability analysis, the integration of Eq.~\eqref{eq: Riccati_outer} is repeated by iterating over the wave speed $\omega/\alpha$ such that the wall-boundary condition is matched. However, we directly use the complex wave speed obtained from LST just to compare the amplified inviscid modes. Comparison in Fig.~\ref{fig: composite_linear} shows that the dynamics of modes may be classified as inviscid, with the exceptions of the wall layer and the critical layer. 

\begin{figure}[!t]
    \centering
    \includegraphics[width=\linewidth]{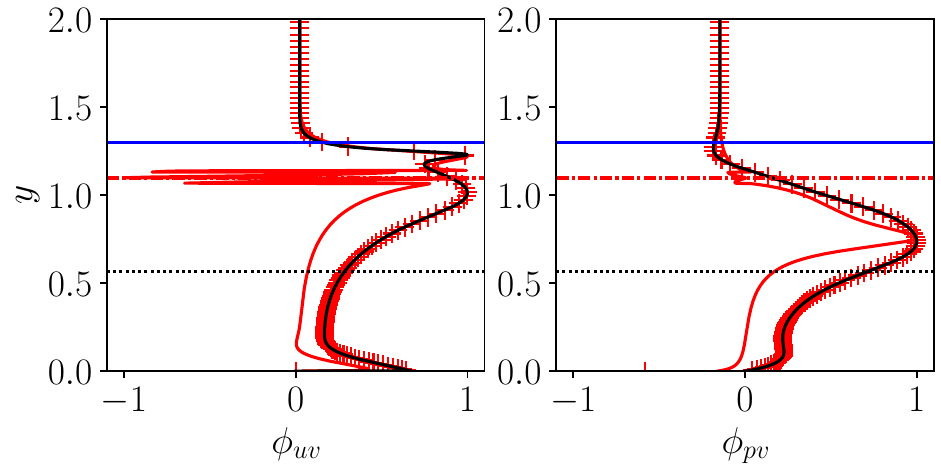}
    \put(-250,125){(a)}
    \put(-125,125){(b)}
    \caption{Comparison of the relative phase $\phi_{uv}$ and $\phi_{pv}$ as obtained from the LST modes (black --), DNS (red $+$), and the asymptotically matched modes using the inviscid  solution and viscous wall-layer solution from Eqs.~\eqref{eq: wall_layer_classic} for M1 case.}
    \label{fig: composite_LST_DNS_phase}
\end{figure}

\begin{figure}[!b]
    \centering
    \includegraphics[width=\linewidth]{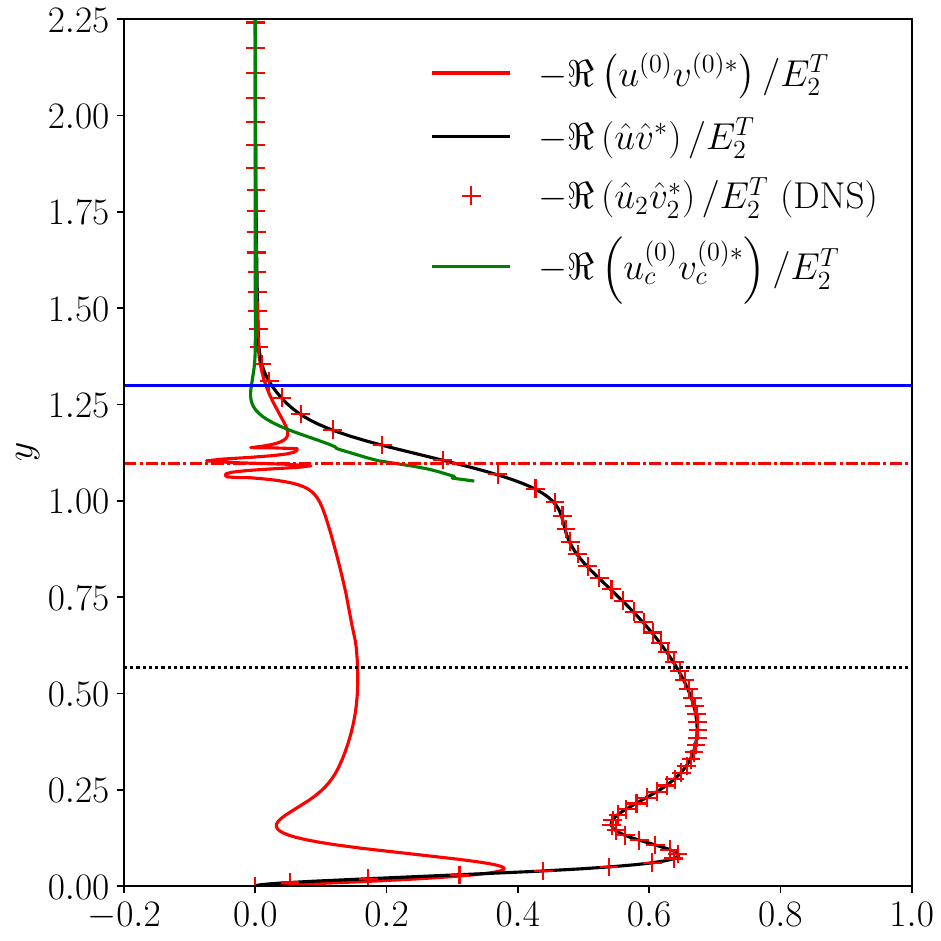}
    \caption{Comparison of the production term (see Eq.~\eqref{eq: pertEnergyLinear}) parameter $-\Re\left(\hat{u}\hat{v}^*\right)$ as obtained by LST against the inviscid analysis for M1 case. The inviscid analysis captures positive growth since the full complex frequency $\omega$ obtained from LST is used in the inviscid analysis.}
    \label{fig: LST_inviscid_production}
\end{figure}

Figure~\ref{fig: composite_LST_DNS_phase} compares the phase difference between $\hat{u}~\&~\hat{v}$ $(\phi_{uv})$ and $\hat{p}~\&~\hat{v}$ $(\phi_{pv})$, defined as (for some variables $a$ and $b$), 
\begin{equation}
    \phi_{ab} = -\frac{\Re\left(\hat{a}\hat{b}^*\right)}{|\hat{a}||\hat{b}|},
    \label{eq: phase_def}
\end{equation}
as obtained from the LST, DNS, and the asymptotically reconstructed modes.  From the asymptotically reconstructed modes, $\hat{p}$ and $\hat{v}$ are primarily out of phase below the relative sonic line ($\phi_{pv}\approx 0$), indicating the inviscid perturbations are standing waves~\cite{kinsler2000fundamentals}. Comparing with the production term in Eq.~\eqref{eq: pertEnergyLinear}, we note that $\phi_{uv}$ captures the production of perturbation energy. The inviscid analysis under-predicts the energy production below the critical line, indicating that below the critical layer, the energy production is primarily due to the viscosity-mediated phase difference between $\hat{u}$ and $\hat{v}$. This is further highlighted in Fig.~\ref{fig: LST_inviscid_production}, which shows that the energy production predicted from the LST (which matches DNS in Fig.~\ref{fig: energy-balance-perturbation}) is significantly higher than the predicted production from the inviscid analysis. Between the critical line and the relative sonic line, pressure perturbations and $y$ direction velocity perturbations have opposite relative phase, indicating the evanescent waves being generated due to the production in the critical layer. While the inviscid waves above the critical layer are very weak (see Fig.~\ref{fig: LST_inviscid_production}), the phasing of the waves is captured correctly (phase difference $>\pi/2$). 

The above discussion highlights the insufficiency of the inviscid perturbations in capturing production. To account for the viscosity within the critical layer of thickness $\delta_c$ of order $(\mathrm{Re}_{\delta^{\star}})^{-1/3} = \epsilon^{2/3}$, a scaled coordinate of the form,
\begin{equation}
    y = y_c + \delta_c\zeta,
    \label{eq: critical_layer_coordinate}
\end{equation}
and the perturbation expansions of the form $\hat{\phi} = \phi^{(0)}_c + \delta_c \phi^{(0)}_c + \mathcal{O}(\delta^2_c, \epsilon)$ are useful~\cite{lees1946investigation, schlichting1950amplitude, reshotko1960stability, haberman1972critical, wu2019nonlinear}. Expanding $f$ and $g$ as 
\begin{align}
 &f = \Re{(c)} + f'_c\delta_c \zeta + \frac{f''_c}{2}\delta^2_c \zeta^2+\cdots\\
 &g = g_c + g'_c\delta_c \zeta + \frac{g''_c}{2}\delta^2_c \zeta^2 + \cdots,
\end{align}
we obtain the following continuity equation,
\begin{equation}
 D_{\zeta}v^{(0)}_c + \delta_c\left(i\alpha u^{(0)}_c + D_{\zeta}v^{(1)}_c\right) + \mathcal{O}(\delta^2_c) = 0.
 \label{eq: critical_layer_continuity}
\end{equation}
Where $D_{\zeta}=\frac{d}{d \zeta}$. There are primarily two methods that follow after the introduced scaling. As discussed by Lees and Lin~\cite{lees1946investigation} and Schlichting~\cite{schlichting1950amplitude}, the full leading order viscous solution can be obtained. However, historically, this used to be solved as a boundary value problem and satisfying $y\to\infty$ and $y=0$ boundary conditions. It is clear from the structure of the eigenmodes that when the critical layer is sufficiently far away from the wall, the $y=0$ boundary condition should be satisfied by the wall-layer solution and not the viscous solution in the critical layer. The critical layer and the wall layer solutions may be connected using WKB approximation solutions of the inviscid problem between the critical line and the wall~\cite{bower1994analytical}. Otherwise, as discussed by Schlichting~\cite{schlichting1950amplitude} and Reshotko~\cite{reshotko1960stability}, the leading order solution may be taken as an inviscid solution with higher order solutions as thermoviscous corrections, although, the singularity in the inviscid solution in the critical layer is only presumed to be eliminated completely by the viscous correction (see appendix G of Reshotko~\cite{reshotko1960stability}). Here, we follow the simpler first method since the aim is to only demonstrate that thermoviscous effects are important for capturing the correct perturbation energy production in the critical layer. 
Equation~\eqref{eq: critical_layer_continuity} yields $v^{(0)}_c = \mathrm{const.}$ and $i\alpha u^{(0)}_c + D_{\zeta}v^{(1)}_c=0$. Using these, we obtain the temperature equation as, 
\begin{align}
 &v^{(0)}_cg'_c + \delta_c\left(i\alpha f'_c \zeta T^{(0)}_c + v^{(0)}_c g''_c\zeta + v^{(1)}_cg'_c - \frac{\gamma g_c}{\mathrm{Pr}}\mu_c D_{\zeta}^2T^{(0)}_c\right) \nonumber \\ &+ \mathcal{O}(\delta^2_c) = 0,
\end{align}
which yields, along with $v^{(0)}_c=0$, the temperature perturbation equation as
\begin{equation}
 i\alpha f'_c \zeta T^{(0)}_c + v^{(1)}_cg'_c = \frac{\gamma g_c}{\mathrm{Pr}}\mu_c D_{\zeta}^2T^{(0)}_c.
\end{equation}
Similarly, the momentum equations yield,
\begin{align}
 &D_{\zeta} p^{(0)}_c + \delta_c D_{\zeta} p^{(1)}_c + \mathcal{O}(\delta^2_c) = 0,\\
 &\delta_c\left(i\alpha f'_c \zeta u^{(0)}_c + v^{(1)}_c f'_c + \frac{i\alpha p^{(1)}_c g_c}{\gamma M^2_\infty}- g_c\mu_c D_{\zeta}^2u^{(0)}_c\right) + \mathcal{O}(\delta^2_c) = 0.
\end{align}
Differentiating the $u^{(0)}_c$ equation, we obtain the following equations for $u^{(0)}_c$ and $T^{(0)}_c$, 
\begin{align}
 i\alpha f'_c \zeta T^{(0)}_c + v^{(1)}_cg'_c = \frac{\gamma g_c \mu_c}{\mathrm{Pr}}D_{\zeta}^2T^{(0)}_c,\label{eq: critical_layer_temperature}\\
 i\alpha f'_c \zeta D_{\zeta}u^{(0)}_c= g_c\mu_c D_{\zeta}^3 u^{(0)}_c.
\end{align}
Solving the momentum equation, we obtain the following (applying the $y\to\infty$ boundary condition), 
\begin{equation}
    u^{(0)}_c = \hat{u}_c\int^\zeta_{\infty}Ai\left((i\beta)^{1/3}\tau\right)d\tau,
\label{eq: viscous_critical_layer}
\end{equation}
where $\beta=\alpha f'_c/g_c\mu_c$ and we use the value of $\hat{u}$ at the critical line obtained from LST. Similarly, Eq.~\eqref{eq: critical_layer_temperature} can be solved upon evaluating $v^{(1)}_c$ using the relation $i\alpha u^{(0)}_c + D_{\zeta}v^{(1)}_c=0$. In Fig.~\ref{fig: LST_inviscid_production}, we also show the variation of the perturbation energy production as obtained using Eq.~\eqref{eq: viscous_critical_layer} and the inviscid solution for $\hat{v}$. Clearly, the variation obtained from the viscous solution follows the one obtained from LST more closely than the inviscid solution, highlighting the importance of the viscous effects in the critical layer for capturing the perturbation energy production in an amplified mode. In the derivation of Eq.~\eqref{eq: viscous_critical_layer} we have ignored the small imaginary part of the wave speed $c$ while deriving Eq.~\eqref{eq: viscous_critical_layer}, ($\alpha \Im\left({c}\right) \ll \delta_c$). 

\begin{figure*}[!t]
    \centering
    \includegraphics[width=\linewidth]{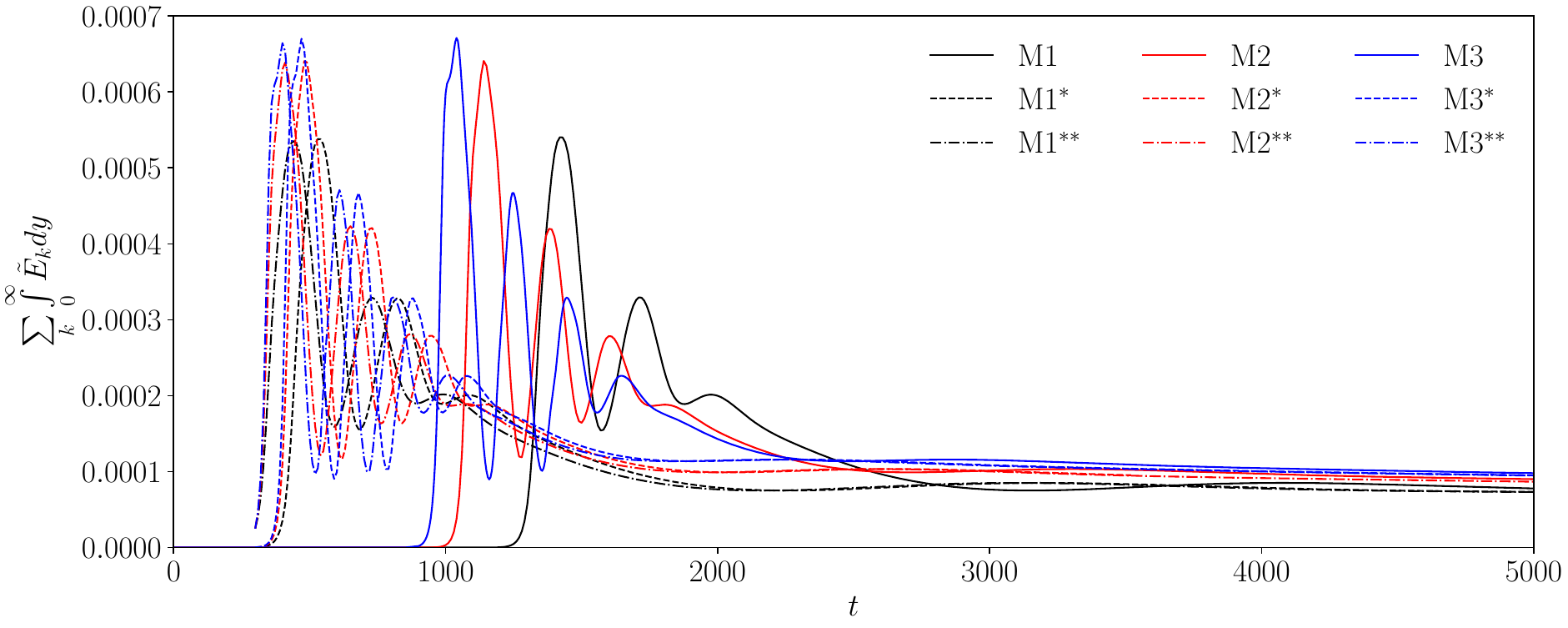}
    \caption{Time series of total perturbation energy for all cases in Table~\ref{tab: parameters_table}. The steady-state energy values do not change based on initial conditions, indicating that the limit cycle reached at a long time is stable.}
    \label{fig: energy_time_series_all}
\end{figure*}

In the next section, we show that the relative phase $\phi_{uv}$ changes drastically between the critical line and the boundary layer edge as the nonlinear saturation takes place, resulting in negative values of the production term in the critical layer. We also show that this change in the phase $\phi_{uv}$ and hence the production is due to the viscous effects in the critical layer in the modes of the distorted base state.

\subsection{Nonlinear regime}
\begin{figure}[!t]
    \centering
    \includegraphics[width=\linewidth]{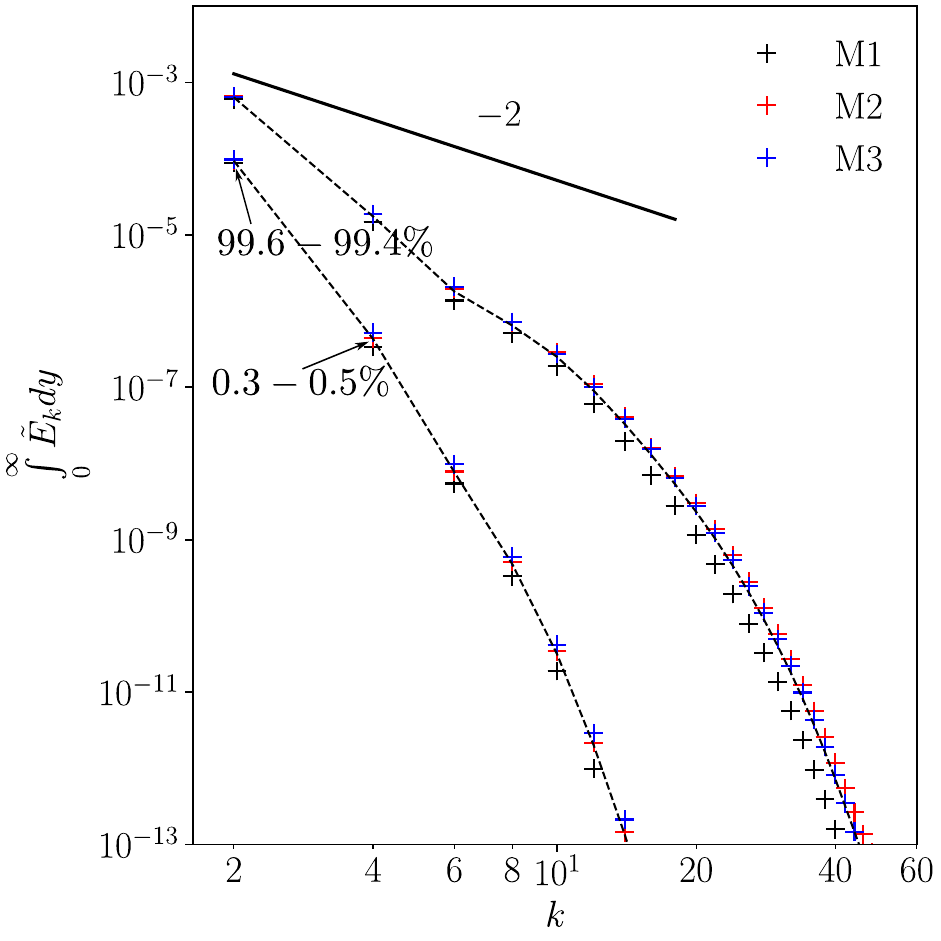}
    \put(-170, 120){Steady}
    \put(-80, 140){Peak}
    \caption{Energy spectra for all three Mach numbers at peak and non-linear saturation stage. The dashed lines are sketched to differentiate the two regimes (peak and steady state). The spectra show that most of the perturbation energy ( $>99\% $) is in $k=2$ mode at steady state. Although sharp structures are visible at peak (see Fig.~\ref{fig:perturbation-energy-semilog}), the spectrum shows that the energy is not sufficient to form the shock waves. \Revision{}{The maximum amplitudes of perturbations for M1 case are as follows (velocity perturbations are relative to freestream streamwise velocity and pressure perturbations are relative to freestream pressure): Peak ($k=2$): $x$ velocity - 5.2\%, $y $ velocity - 4.0\%, pressure - 21\%; Peak ($k=4$): $x$ velocity - 0.8\%, $y$ velocity - 0.57\%, pressure - 4.8\%; Steady ($k=2$): $x$ velocity - 1.8\%, $y$ velocity - 1.3\%, pressure - 7.38\%; Steady ($k=4$): $x$ velocity - 0.12\%, $y$ velocity - 0.06\%, pressure - 0.63\%. }}
    \label{fig: energy_spectra}
\end{figure}

As the amplitude of the unstable mode increases due to the energy production $\mathbb{P}$ (Eq. \eqref{eq: pertEnergyLinear}), it cascades its energy into higher harmonics as a nonlinear effect~\cite{gupta2017spectral, gupta2018spectral}. \Revision{}{This energy cascade takes place via triadic terms in the nonlinear governing equations (see appendix~\ref{appendix: nonlinear_energy}).} Wavenumber of each of these harmonics is an integer multiple of that of the unstable mode. The shape of these higher harmonics in the wall-normal direction is different from the Mack's higher modes (greater than the second mode predicted by LST~\cite{mack1969boundary}). An additional nonlinear effect is the mean flow distortion which results from the self-interaction of the high amplitude unstable mode~\cite{stuart1958non}. In this section, we discuss these nonlinear effects in detail. We show that the higher harmonic generation is coherent. Additionally, we show that the mean flow distortion results in a distortion of the perturbation energy production, which results in the nonlinear \Revision{energy saturation}{saturation of the perturbation energy}. This mechanism is different from the predicted mechanism of saturation in hydrodynamic instability, first discussed by \citeauthor{stuart1958non}~\cite{stuart1958non}. \Revision{}{We use linear stability theory on the distorted base state and the asymptotic results considered in the previous section to highlight the modification in the perturbation energy growth. As shown in Fig.~\ref{fig: energy_spectra}, the relative values of maximum perturbation amplitudes are approximately two orders of magnitude smaller than the freestream values. Hence linear stability calculations on the distorted base state are sufficiently accurate.}
\subsubsection{Higher harmonic generation}
As the instability modes get amplified, higher harmonics are generated~\cite{gupta2017spectral, gupta2018spectral} due to nonlinear steepening. The instability mode gains energy, which is lost to dissipation through higher harmonics. This results in a peak of the mode amplitude before saturation. At the peak, the perturbation energy spectra are widest (see Fig.~\ref{fig: energy_spectra}). Unlike the classical planar steepening, the waves exhibit a more complex structure. As visible in Fig.~\ref{fig:perturbation-energy-semilog}, the waves develop sharp structures, particularly in the regions of highly negative dilatation, indicating nonlinear wave steepening. These sharp waves interact with the walls, modulating the shear and heat flux~\cite{sahithi2024thermoviscous}. Usually, these nonlinear waves continuously steepen till the formation of shock waves, which dissipate energy at the shock thickness scale~\cite{jossy2023baroclinic}. However, the perturbation energy in the unstable mode is not enough for the cascade to result in shock waves~\cite{gupta2017spectral, gupta2018spectral}. After the peak, the energy decreases, and then, after a few oscillations, saturates to a finite value. To extract higher harmonics, we assume the following decomposition of perturbation variables $(u', v', p')$
\begin{equation}
\left(u', v', p'\right) = \sum_k \left(\tilde{u}_k,\tilde{v}_k,\tilde{p}_k\right)(y,t)e^{i\alpha_k x} + \mathrm{c.c},
\label{eq: higher-harmonics-1}
\end{equation}
where $k=2$ denotes the unstable Mack mode and $k=2n, n\geq2$ denotes higher harmonics of the unstable Mack mode. In various nonlinear regimes, the perturbation energy can be calculated using $\tilde{u}_k, \tilde{v}_k, \tilde{p}_k$ obtained from FFT in $x$ direction at different $y$ locations. As shown in Fig.~\ref{fig: energy_spectra}, at the saturated steady state, the energy is significantly lower in all higher harmonics of the unstable mode (less than $0.5\%$ in $k\geq 4$). 

\begin{figure}[!b]
    \centering
    \includegraphics[width=\linewidth]{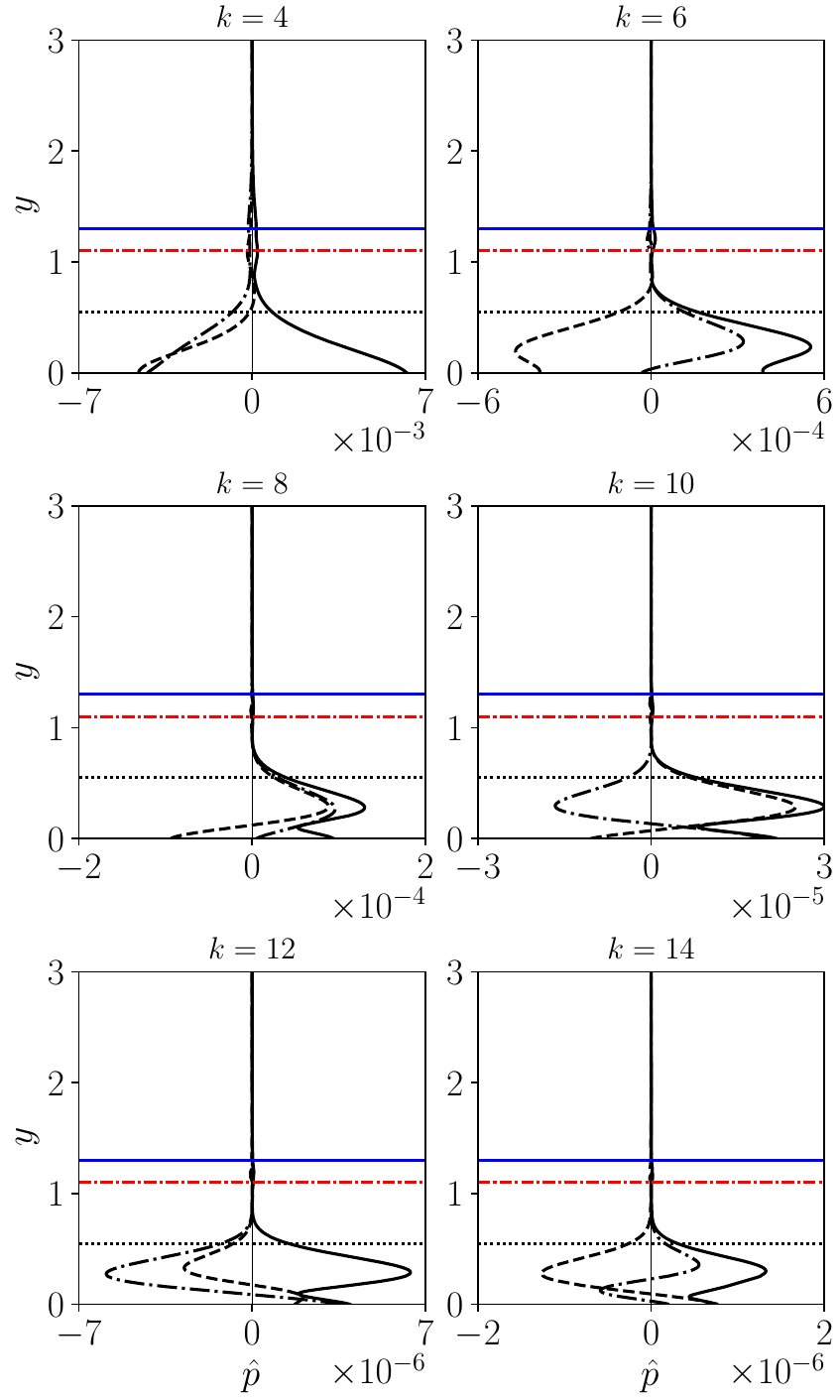}
    \put(-240,410){(a)}
    \put(-120,410){(b)}
    \put(-240,275){(c)}
    \put(-120,275){(d)}
    \put(-240,135){(e)}
    \put(-120,135){(f)}
    \caption{Mode shapes (pressure perturbations) of higher harmonics of the unstable Mack mode in the nonlinear regime at steady state for the M1 case. The amplitude decreases as $k$ increases significantly. Furthermore, only $k=4$ and $k=6$ exhibit some perturbations past the relative sonic line (turning point). Most of the other modes decay exponentially past the relative sonic line.}
    \label{fig: higher_harmonics_shape}. 
\end{figure}

\begin{figure}[!t]
        \centering
        \includegraphics[width=\linewidth]{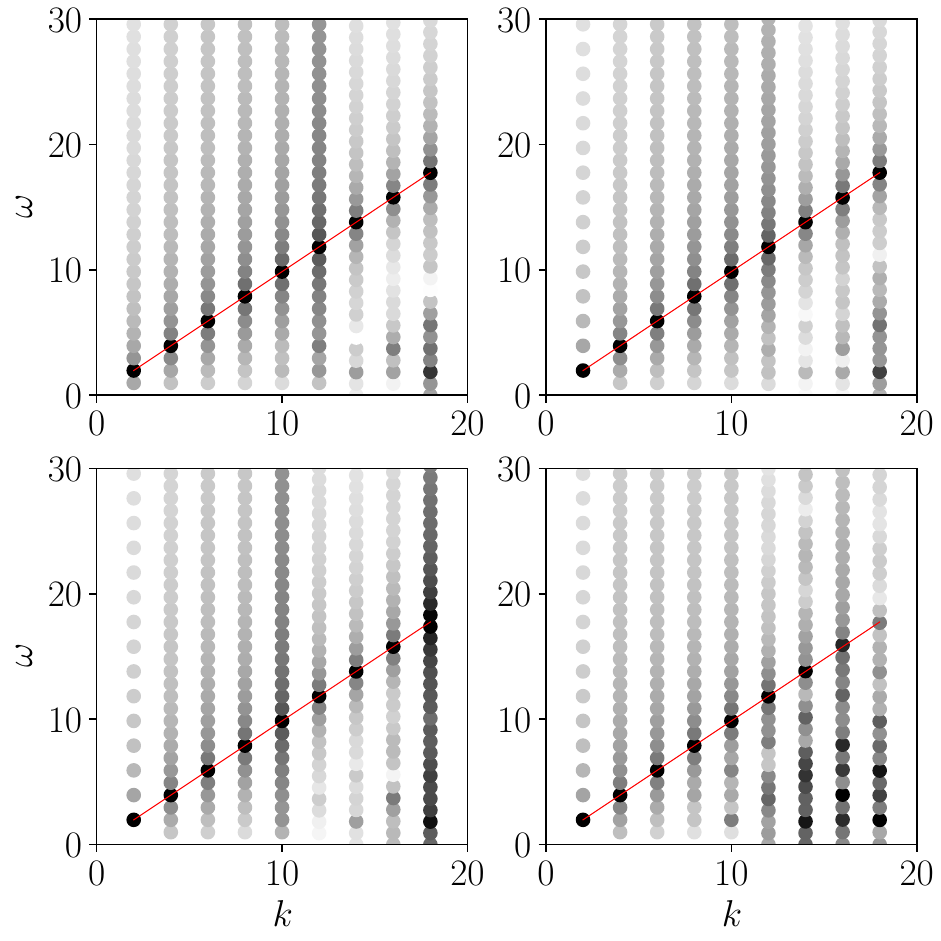}
        \put(-250,250){(a)}
        \put(-120,250){(b)}
        \put(-250,130){(c)}
        \put(-120,130){(d)}
        \caption{Dispersion relation at steady state a) near wall ($y=0.1$), b) near sonic line $f =\frac{\sqrt{g}}{M_{\infty}}$, c) near relative sonic line, and d) near the critical line. For higher harmonics, the dominant frequency is a multiple of the frequency of the unstable mode, with the multiplying factor being the same for $\omega$ and $k$. For higher harmonics, since the signal is weak, the ratio of energy in the dominant frequency to the other frequencies decreases.}
        \label{fig: dispersion-relation}
    \end{figure}

In Fig.~\ref{fig: higher_harmonics_shape} we show pressure perturbation variation in $y$ for $k=4, 6, 8 \cdots 14$ at dimensionless time $t=9700$. The pressure amplitude decreases by orders of magnitude for higher harmonics, as also evident in Fig.~\ref{fig: energy_spectra}. Pressure perturbations above the relative sonic line are significant only for $k=4$ and $k=6$, highlighting that most of the higher harmonic generation takes place below the relative sonic line (the classical acoustic turning point). Beyond the turning point, the waves become evanescent with only a small amplitude near the critical layer. Equation~\eqref{eq: higher-harmonics-1} can be further decomposed (for pressure) as $\tilde{p}_k = \hat{p}_k(y)e^{-i\alpha_k c_k t}$ where $c_k$ is the phase speed for higher harmonics. In Fig.~\ref{fig: dispersion-relation}, we show the dispersion relation $\omega_k$ vs $k$ for higher harmonics. The dark markers highlight the peak frequency obtained in the signal after an FFT operation in time. We calculate this relation at a few important $y$ locations, namely near wall ($y=0.1$), near sonic line ($f = \sqrt{g}/M_\infty$), near relative sonic line, and near critical line. \RevisionTwo{Frequency $\omega_k$} increases proportionately with $k$, and the phase speed $c_k$ remains almost constant. This shows that the higher harmonics are generated due to propagating non-planar finite amplitude unstable mode without any dispersion or separation of modes. Additionally, \RevisionTwo{the higher harmonics} differ from the traditional higher modes obtained from the inviscid Rayleigh Eq.~\eqref{eq: pridmore-brown-outer}~\cite{mack1969boundary}.

The nonlinear steepening is simply an artifact of high amplitude in the instability mode. The higher harmonics generated due to the nonlinear steepening of the unstable Mack modes ride with the modes with equal phase speed. Since the contribution of the modes to the total perturbation energy is insignificant (c.f.~Fig.~\ref{fig: energy_spectra}), these modes have an insignificant effect on the dynamics. As discussed in the next section, high amplitude in the unstable mode contributes to the mean-flow distortion. This mean flow distortion further changes the unstable mode phasing, thus reducing the net perturbation energy production which eventually results in the saturation of the perturbation energy.

\subsubsection{Mean flow and mode distortion}
\label{sec: mean-flow-distortion}
The perturbation energy production term in Eq.~\eqref{eq: pertEnergyLinear} comprises a Reynolds stress type term $\Re\left(\hat{u}\hat{v}^*\right)$ multiplied with the mean flow density $1/g$ and shear $df/dy$. As the nonlinear propagation of the unstable mode continues, the self-interaction of the mode results in mean flow distortion, similar to acoustic streaming~\cite{lighthill1978acoustic}. As a result, the mean flow variables $f$ and $g$ get distorted. As a result of this distortion, the energy production changes.~\citeauthor{stuart1958non}~\cite{stuart1958non} introduced the concept of nonlinear saturation of near-neutral instability modes due to mean flow distortion. However, in his analysis, the Reynolds stress type term was considered to vary only due to the amplitude of the perturbations with constant shape. Based on Stuart's analysis, several hydrodynamic instability theories have been developed to understand the nonlinear saturation of the unstable modes through a nonlinear amplitude equation~\cite{ stuart1960non, watson1960non, stewartson1971non, wu2019nonlinear}. Assuming the mean flow gets distorted due to quadratic self-interactions of the high-amplitude unstable mode, incorporating these in the energy production terms results in a quadratic nonlinear time evolution equation for the amplitude square, which is a measure of the perturbation energy. In this section, we show that the modification in the Reynolds stress term of the perturbation energy dominates the mean flow distortion as the nonlinear regime progresses. This distortion in the Reynolds stress term results in a drop of the total perturbation energy production just enough to cancel the dissipation, resulting in a steady state of the flow. 

\begin{figure}[!t]
    \centering
    \includegraphics[width=\linewidth]{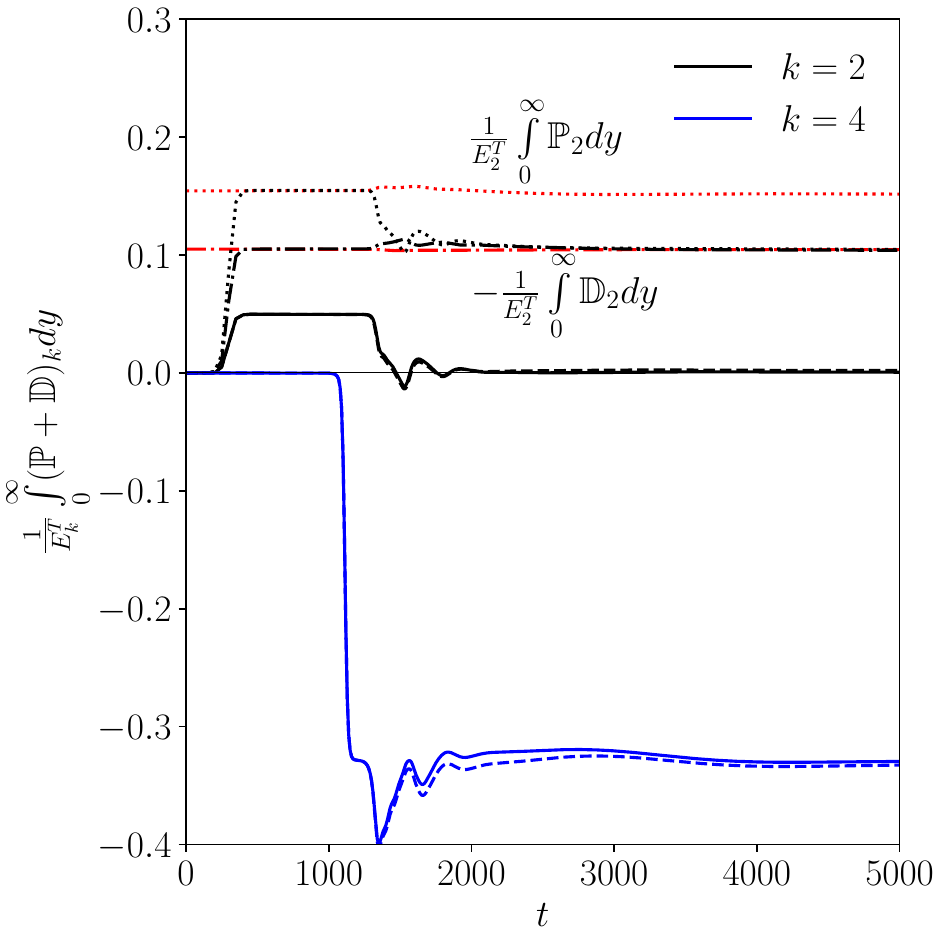}
    \caption{Net linear production of the modes for M1 case (solid line `-': using instantaneous modes and instantaneous mean flow; dashed line `- -' : using instantaneous modes and similarity solution) scaled with instantaneous energy of the respective modes. The chained line shows dissipation of the unstable mode ($k=2$) (red chained {\color{red} `- $\cdot$'}: linear regime mode on instantaneous mean flow; black chained {\color{black} `- $\cdot$'} : instantaneous mode on instantaneous mean flow). The dotted line `$\cdots$' shows production of
    the unstable mode ($k=2$) (red dotted {\color{red} `$\cdots$'} : linear regime modes on instantaneous mean flow; black dotted {\color{black} `$\cdots$'} : instantaneous modes on the instantaneous mean flow). }
    \label{fig: net-production}
\end{figure}

To model the mean flow distortion, we consider the decomposition of the fields as
\begin{subequations} \label{eq: decomposition-distortion}
\begin{align}
      &\rho(x,y, t) = \underbrace{\frac{1}{g(y)}+ \epsilon^2_a\rho_D(y,t)}_{\overline{\rho}(y, t)}  + \epsilon_a \rho'(x,y,t) ,\\
      &p(x,y,t) = \underbrace{1 + \epsilon^2_a p_D(y,t)}_{\overline{p}(y,t)} + \epsilon_a p',\\
      &T(x,y, t) = \underbrace{g(y)+ \epsilon^2_a T_D(y,t)}_{\overline{T}(y,t)} + \epsilon_a T'(x,y,t) ,\\
    &u(x,y, t) = \underbrace{f(y)+ \epsilon^2_a u_D(y,t)}_{\overline{u}(y,t)} + \epsilon_a u'(x,y,t) , \\
    &v(x,y,t) = \underbrace{\epsilon^2_a v_D(y,t)}_{\overline{v}(y,t)} + \epsilon_a v'(x,y,t).
\end{align}
\end{subequations}
where the subscript $()_D$ denotes distortion in mean flow and $\epsilon_a$ is amplitude parameter (see Eq.~\eqref{eq: amplitude_decomp}). The $\overline{\left(\right)}$ denotes averaging in $x$ over a single wavelength $2\pi/\alpha$. Here, we consider the distortions are uniformly applied in $x$, since they happen over a time scale significantly longer than the time scale of the unstable mode propagation. From DNS, the instantaneous distorted mean flow can be obtained using the $k=0$ component of FFT in $x$ at every $y$ location.  
In Fig.~\ref{fig: net-production}, we show the net production due to base flow interactions (R.H.S. of Eq.~\eqref{eq: pertEnergyLinear}), production, and negative of dissipation normalized by the instantaneous perturbation energy using the instantaneous perturbation modes for $k=2$ and $k=4$ based on the similarity solution base state ($f$ and $g$) and the instantaneous distorted based state ($\overline{u}$ and $\overline{T}$). We also plot the production and the negative of dissipation terms, assuming the shape of the perturbation modes remains constant (identical to those in the linear regime). \RevisionTwo{We plot} only the production due to base flow interactions. The net production of each mode in the nonlinear regime will consist of an energy exchange term as well (see derivation in appendix~\ref{appendix: nonlinear_energy}). At the steady state, the sum of net production due to the interaction with the base state and the nonlinear exchange term is zero.
\RevisionTwo{Replacing $f$ and $g$} with $\overline{u}$ and $\overline{T}$, respectively, makes no significant change in the net production of the perturbation energy computed using the instantaneous mode shapes. Moreover, for $k=4$, the net production \Revision{}{due to mean flow interaction} based on Eq.~\eqref{eq: pertEnergyLinear} is always negative. This implies that the mode obtains energy from $k=2$ due to nonlinear cascade, which is quickly dissipated by viscosity and mean flow interaction. For $k=2$, the normalized net production reaches a maximum and remains almost constant in the linear regime. However, with increasing energy in $k=2$, the mean flow begins to distort (along with cascade in higher harmonics). This mean flow distortion changes the shape of the characteristic mode shape on the new base state (distorted). This change in the mode shape results in a significant decrease in the net production till it approaches zero, signifying a steady state. The decrease in the net production and only production term are very similar with only minor changes in the dissipation. Importantly, the production and the dissipation terms evaluated assuming constant mode shapes (identical to those in the linear regime) change almost insignificantly throughout the linear and nonlinear regimes. This suggests that the mechanism proposed by Stuart~\cite{stuart1958non} for nonlinear saturation of the modes in incompressible shear flows does not apply to the nonlinear saturation of the Mack modes. This can be further highlighted considering the following model formulation. 

Performing $x$ averaging of the dimensionless and conservative form of Eqs.~\eqref{eq: Navier-Stokes dimensional} we obtain,
\begin{subequations}
   \begin{align}
       &\frac{\partial \overline{\rho}}{\partial t} + \frac{\partial \overline{\rho v}}{\partial y} = 0,\\
       &\frac{\partial \overline{\rho u}}{\partial t} + \frac{\partial \overline{\rho u v}}{\partial y} = \frac{1}{\mathrm{Re}_{\delta^{\star}}}\frac{\partial}{\partial y}\left(\overline{\mu\frac{\partial u}{\partial y}} - \mu_0\frac{df}{dy}\right),\\
       &\frac{\partial \overline{\rho v}}{\partial t} + \frac{\partial \overline{\rho v^2}}{\partial y} + \frac{1}{\gamma M^2_\infty}\frac{\partial \overline{p}}{\partial y}=0,\\
       &\frac{\partial }{\partial t}\left(\overline{\rho T} + \frac{\gamma(\gamma - 1) M^2_\infty}{2}\overline{\rho|\boldsymbol{u}|^2}\right) + \frac{\partial}{\partial y}\left(\overline{\rho v T}\right.\nonumber\\&\left. + \frac{\gamma(\gamma - 1) M^2_\infty}{2}\overline{\rho v |\boldsymbol{u}|^2} + (\gamma - 1)\overline{p v}\right) = \frac{\gamma }{\mathrm{Re}_{\delta^{\star}}\mathrm{Pr}}\frac{\partial}{\partial y}\left(\overline{\mu \frac{\partial T}{\partial y}}\right.\nonumber\\&\left.-\mu_0\frac{dg}{dy}\right)+\frac{M^2_\infty\gamma(\gamma -1)}{\mathrm{Re}_{\delta^{\star}}}\frac{\partial}{\partial y}\left(\overline{\mu u \frac{\partial u}{\partial y}} - \mu_0 f\frac{df}{dy}\right).
   \end{align} 
\label{eq: distortion_gov}
\end{subequations}
In Eqs.~\eqref{eq: distortion_gov}, we have also applied the boundary layer scaling to the governing equations while considering the dimensionless form. Consequently, viscous effects in the $y$ momentum equation and due to streamwise gradients in $x$ momentum and total energy equation drop out. 
Substituting the decompositions in Eqs.~\eqref{eq: decomposition-distortion} in Eqs.~\eqref{eq: distortion_gov}, and noting that the quantities are periodic in $x$, we get the following expressions for the distorted mean flow variables $\overline{v}$, $\overline{p}$, and $\overline{\rho}$ 
    \begin{equation}
    \overline{v} = - g\overline{\rho' v'},~~\overline{p} = 1-\frac{\gamma M_{\infty}^2 \overline{v'^2}}{g},~~\overline{\rho} = \frac{\overline{p}- \overline{\rho' T'}}{\overline{T}}
    \end{equation}
    and $\overline{u}$ and $\overline{T}$ at the steady state,
    \begin{subequations}\label{eq: distorted-mean-flow}
    \begin{align}
    &\frac{d\overline{u}}{dy} =\frac{1}{\overline{\mu}}\left(\mu_0 \frac{df}{dy} - \overline{\mu'\frac{\partial u'}{\partial y}} + \mathrm{Re}_{\delta^{\star}} \frac{\overline{u'v'}}{g}\right),\label{subeq: distorted_u} \\
    &\frac{d\overline{T}}{dy} = \frac{1}{\overline{\mu}}\left(\mu_0 \frac{dg}{dy} - \overline{\mu' \frac{\partial T'}{\partial y}} - (\gamma -1)M^2_{\infty}\mathrm{Pr}\left((\overline{u}-f) \overline{\mu\frac{\partial u}{\partial y}}\right. \right.\nonumber \\
    &\left. \left.+\overline{\mu' u'} \frac{df}{dy} + \mu_0 \overline{u' \frac{\partial u'}{\partial y} }\right) + \mathrm{Re}_{\delta^{\star}}\mathrm{Pr} \frac{\overline{v'T'}}{g}\right),\label{subeq: distorted_T}
\end{align}
\end{subequations}
where we have used the condition that far away from the boundary layer, gradients of the distorted mean flow must vanish. Equations~\eqref{eq: distorted-mean-flow} may be compared with the mean flow distortion equations given by~\citeauthor{stuart1958non}~\cite{stuart1958non} for incompressible shear flows. Consequently, we write the energy budget for $k^\mathrm{th}$ mode in the nonlinear regime as, 
\begin{align}
    &\frac{\partial \tilde{E}_k}{\partial t}  + \frac{1}{\gamma M^2_{\infty}}\frac{\partial}{\partial
    y}\Re\left(\tilde{p}_k\tilde{v}^*_k\right) =  -\frac{1}{\overline{T}}\frac{d\overline{u}}{dy}\Re\left(\tilde{u}_k\tilde{v}^*_k\right) + \mathbb{D}_k + \mathcal{N}_k,
    \label{eq: pertEnergyNonLinear}
\end{align}
where $\mathbb{D}_k$ corresponds to the three dissipation terms for the $k^{\mathrm{th}}$ mode, $\mathcal{N}_k$ corresponds to the loss or gain of energy in the $k^{\mathrm{th}}$ mode due to nonlinear interactions, and $\tilde{E}_k$ is the energy in the $k^{\mathrm{th}}$ mode with distorted mean flow, defined as, 
\begin{equation}
    \tilde{E}_k = \frac{1}{2}\left(\frac{|\tilde{u}_k|^2 + |\tilde{v}_k|^2}{\overline{T}} + \frac{|\tilde{p}_k|^2}{(\gamma M_\infty)^2}\right).
    \label{eq: spectral_energy}
\end{equation}
\Revision{}{Derivation of the nonlinear term in Eq.~\eqref{eq: pertEnergyNonLinear} is given in appendix~\ref{appendix: nonlinear_energy}.} Given that $k=2$ mode extracts energy from the base state and cascades it into higher harmonics, $\mathcal{N}_2 < 0$ and $\mathcal{N}_k>0$ for $k>2$. In general, $\mathcal{N}_k$ can be derived by using the $\mathcal{O}\left(\epsilon^2_a\right)$ terms for perturbation variables in Eqs.~\eqref{eq: decomposition-distortion}. In Fig.~\ref{fig: net-production} we show the RHS of Eq.~\eqref{eq: pertEnergyNonLinear} for $k=2$ and $k=4$ without the nonlinear cascade term $\mathcal{N}_k$, normalized by the total energy in the mode $E^T_k = \int\displaylimits^\infty_0\tilde{E}_kdy$. For $k=2$, the quantity saturates at a small positive value ($\approx 5.5\times 10^{-4}$), which is nullified by negative $\mathcal{N}_2$ at steady state. Additionally, for $k=4$, the quantity saturates at a negative value of $\approx-0.33$, which is nullified by positive $\mathcal{N}_4$. The quantity shown in Fig.~\ref{fig: net-production} is normalized with the total perturbation energy in the corresponding mode. Since $E^T_4\ll E^T_2$, the magnitude of the normalized net production due to mean flow interaction for $4^{\mathrm{th}}$ mode is greater than that for $2^{\mathrm{nd}}$ mode. \Revision{}{While \citeauthor{zhu2020nonlinear}\cite{zhu2020nonlinear} have indicated that for a spatially developing hypersonic boundary layer over a flared cone, higher harmonics may be amplified due to mean flow interactions, in the present scenario, the mechanism of generation of higher harmonics is the inter-modal interaction, as highlighted by the triadic interaction terms derived in appendix~\ref{appendix: nonlinear_energy}.}

\begin{figure}[!t]
    \centering
    \includegraphics[width=\linewidth]{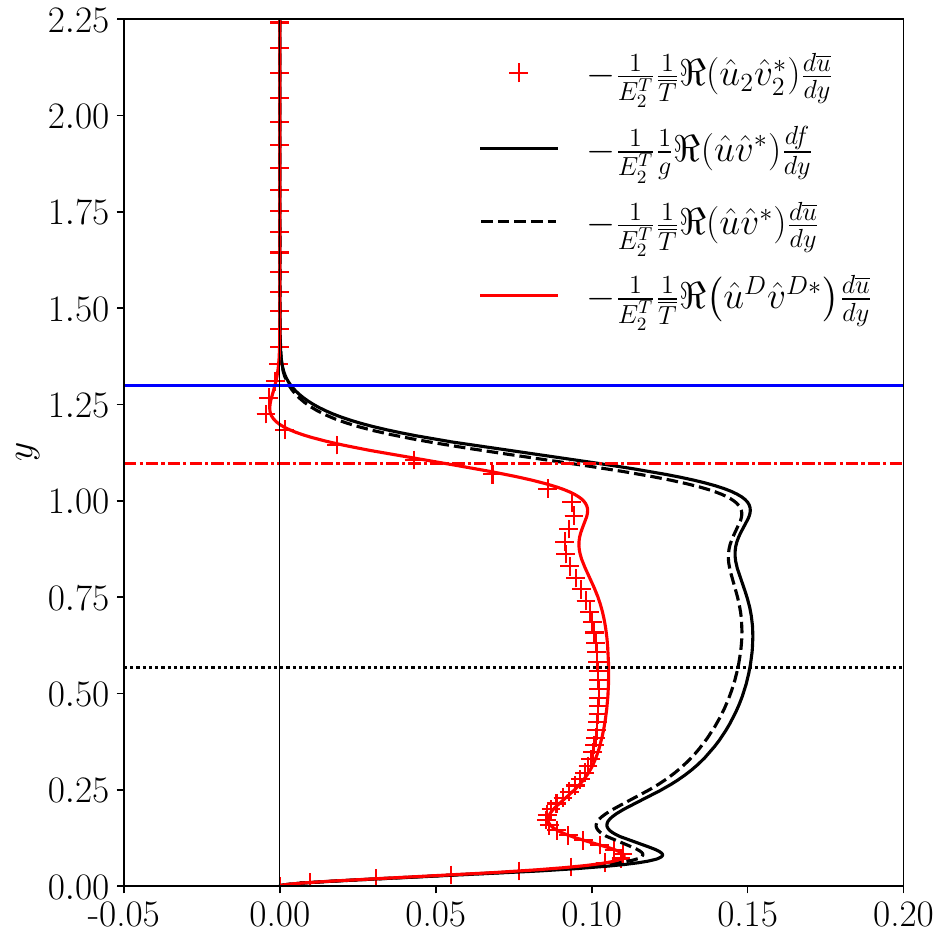}
    \caption{Comparison of normalized production term at steady state calculated using LST modes on distorted mean flow and distorted mean flow gradients, LST modes on laminar mean flow and laminar mean flow gradients, LST modes on laminar mean flow and distorted mean flow gradients, against the value extracted from DNS for M1 case.}
    \label{fig: production_saturation_comparison}
\end{figure}
To further highlight the correct mechanism of saturation of perturbation energy, we distinguish the $k=2$ modes computed using LST on the base state $f$ and $g$ as $\hat{(\cdot)}$ and those computed on the distorted base state $\overline{u}$ and $\overline{g}$ as $\hat{(\cdot)}^D$. The $k=2$ modes extracted from the DNS are denoted by $\hat{(\cdot)}_2$. In Fig.~\ref{fig: production_saturation_comparison}, we compare the four possible normalized production terms. Since~\citeauthor{stuart1958non}\cite{stuart1958non, stuart1960non} and \citeauthor{watson1960non}\cite{watson1960non}, the accepted saturation mechanism in temporally unstable shear flows has been the reduction in energy production calculated using the instability modes on the laminar base state while using the distorted base state in the production term. However, as shown in Fig.~\ref{fig: production_saturation_comparison}, the reduction by just replacing the laminar base state variables $f$ and $g$ with the distorted mean flow variables $\overline{u}$ and $\overline{T}$ is insignificant. However, if we use the distorted base state to compute the modes from LST, the production term follows the values extracted from the DNS. This shows that the modification of the quantity $\Re\left(\hat{u}\hat{v}^*\right)$ due to the mean flow distortion is the primary mechanism resulting in the saturation of the perturbation energy. 

Based on the above discussion, we propose a numerical model for computing the mean flow distortion and the perturbation energy at the steady state. Since more than $99\%$ of the perturbation energy is in $k=2$ mode (which is linearly unstable), we can solve for the steady state perturbation energy in $k=2$ mode based on Eq.~\eqref{eq: pertEnergyNonLinear} as, 
\begin{align}
    &{E}^{T, (m+1)}_2 \nonumber \\ &= {E}^{T, (m)}_2 + h\int\displaylimits^\infty_0 \left(\mathbb{D}^{(m)}-\frac{1}{\overline{T}^{(m)}}\frac{d\overline{u}^{(m)}}{dy}\Re\left(\hat{u}^{(m)}\hat{v}^{*,(m)}\right)\right)dy,
    \label{eq: numerical_model}
\end{align}
 where $(m)$ denotes the iteration number, $h$ denotes a suitable step size, and $\mathbb{D}^{(m)}$ the total viscous dissipation term based on the LST modes evaluated using the distorted flow variables $\overline{u}^{(m)}$ and $\overline{T}^{(m)}$, which are evaluated using the coupled ODEs Eq.~\eqref{eq: distorted-mean-flow}. \RevisionTwo{The production} term is also evaluated using the LST modes on the distorted flow variables. By only using the LST modes evaluated using steady laminar flow variables $f$ and $g$, the perturbation energy is expected to be significantly over-predicted, since the reduction in production is very small (see Fig.~\ref{fig: production_saturation_comparison}). The iterations in Eq.~\eqref{eq: numerical_model} are carried out till the relative difference between $E^{T,(m+1)}_2$ and $E^{T,(m)}_2$ is negligible. Additionally, the updated LST modes may also be evaluated using the sensitivity matrices~\cite{park2019sensitivity} and the base flow distortion due to nonlinearities in Eq.~\eqref{eq: distorted-mean-flow}. 
 \begin{figure}[!t]
    \centering
    \includegraphics[width=\linewidth]{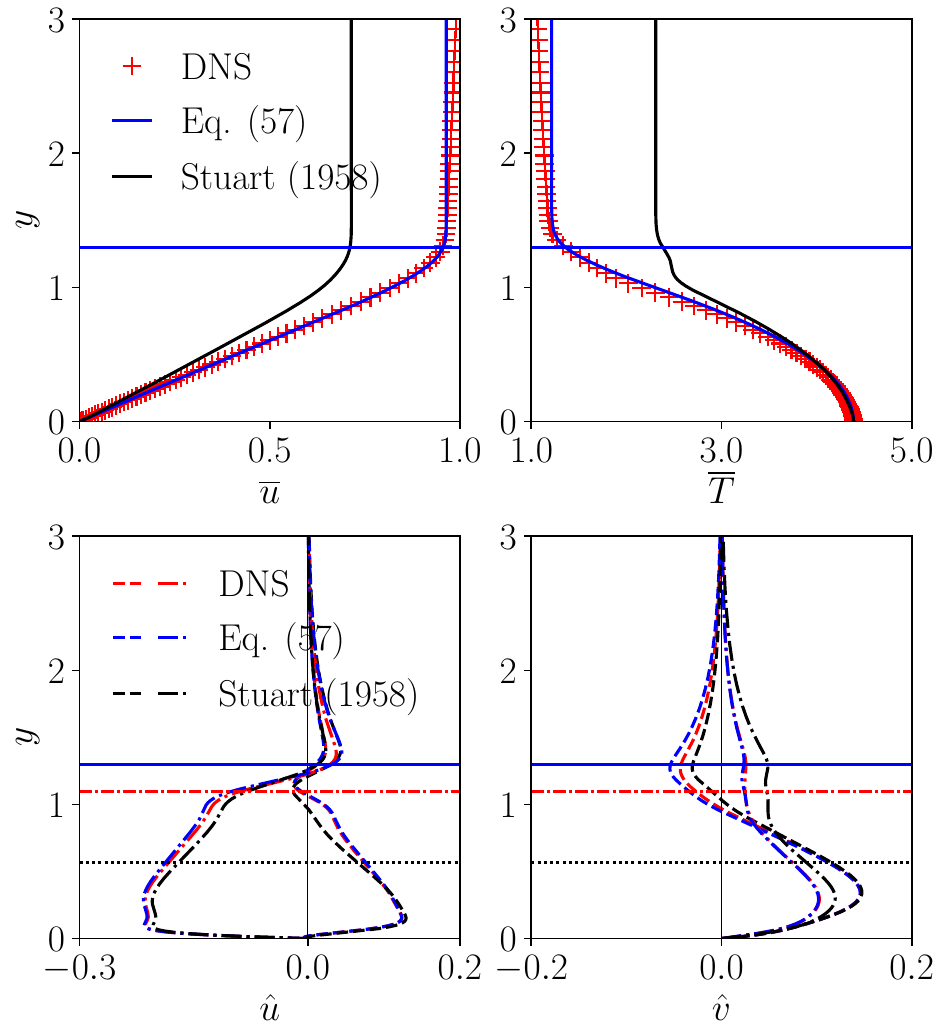}
    \put(-240,275){(a)}
     \put(-120,275){(b)}
      \put(-240,140){(c)}
       \put(-120,140){(d)}
    \caption{Comparison of a) distorted steady mean streamwise velocity, b) distorted steady mean temperature, c) $\hat{u}$, and d) $\hat{v}$ as calculated by current model (Eq. \eqref{eq: numerical_model}) and Stuart model\cite{stuart1958non} with DNS for M1 case. The mode shapes are scaled to yield equal perturbation energy for shape-comparison. The current model accurately captures the distorted steady mean flow and mode shapes.  }
    \label{fig: steady_state_model}
\end{figure}
\begin{figure}[!t]
    \centering
    \includegraphics[width=\linewidth]{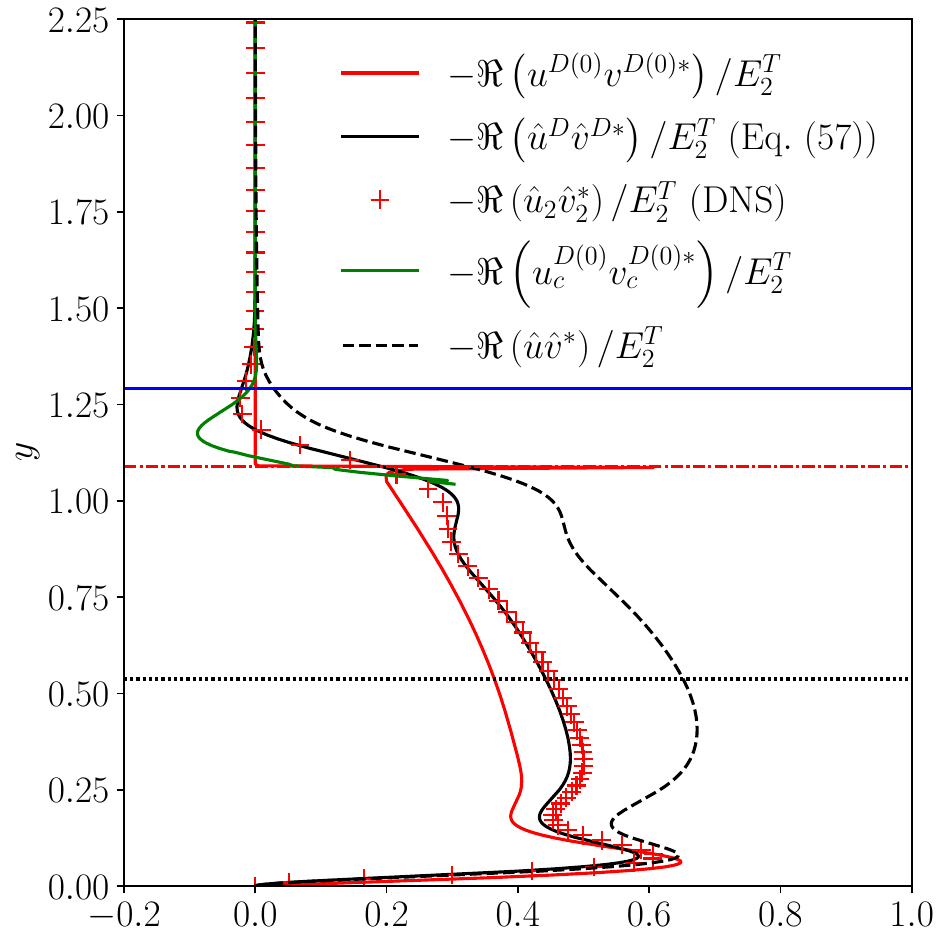}
    \caption{Comparison of normalized Reynolds stress term $-\Re\left(\hat{u}\hat{v}^*\right)$ as calculated using the modes obtained from the model calculation in Eq.~\eqref{eq: numerical_model}, the inviscid solution with wall layer and critical layer viscous corrections, LST modes from laminar mean flow against DNS at steady state for M1 case. }
    \label{fig: ReynoldsStress_Model_comparison}
\end{figure}
We evaluate the steady state distorted mean flow $\overline{u}$ and $\overline{T}$ and the perturbation energy $E^T_2$ using the model Eq.~\eqref{eq: numerical_model}.  In Fig.~\ref{fig: steady_state_model}, we show the distorted mean flow computed from the model in comparison with the DNS. We also evaluate the steady state mean flow and modes assuming no distortion in the modes (~\citeauthor{stuart1958non}~\cite{stuart1958non}). As shown in Fig.~\ref{fig: steady_state_model}, the freestream velocity decreases, and the temperature increases due to the nonlinear distortion, indicating viscous deceleration of the flow. However, the decreased value of freestream velocity is $\approx0.96$ and the increased value of freestream temperature is $\approx1.2$. On the other hand, the distortion predicted upon assuming constant mode shapes~\cite{stuart1958non} are significantly overpredicted. Additionally, $\overline{u}$ and $\overline{T}$ from DNS exhibit small gradients outside the boundary layer. In appendix~\ref{appA: steady_state_DNS}, we show that these residual gradients simply mean that the DNS results are not exactly at the steady state but evolving very slowly. In Fig.~\ref{fig: steady_state_model}c, d, we show the modes obtained at the steady state of calculation from Eq.~\eqref{eq: numerical_model}, DNS, and the calculation from Eq.~\eqref{eq: numerical_model} assuming constant mode shapes. We have normalized the energy for all the modes shown for comparison of the mode shapes. In Fig.~\ref{fig: ReynoldsStress_Model_comparison}, we show that the normalized Reynolds stress term $\Re\left(\hat{u}\hat{v}^*\right)$ responsible for production decreases in the critical layer as well as below the critical layer. The inviscid modes $u^{(0)}$ and $v^{(0)}$ with wall layer correction and the viscous corrected critical layer modes $u^{D(0)}_c$ and $v^{D(0)}_c$ are calculated using the distorted mean flow obtained from the model Eq.~\eqref{eq: numerical_model}. \RevisionTwo{The normalized} Reynolds stress obtained from the model and that from the saturated base state from DNS match very well. Moreover, the overall variation is captured well by the inviscid modes with viscous corrections in the wall layer and the critical layer. Moreover, without assuming constant mode shapes, the relative phasing between streamwise and wall-normal velocity perturbations is captured correctly, particularly within the critical layer. On the other hand, assuming constant mode shapes results in continuous growth of perturbation energy till the mean flow distortion is enough to reduce the production. In Table.~\ref{tab: energy_compare_model}, we compare the steady state perturbation energy values obtained from the Eq.~\eqref{eq: numerical_model}, DNS, and calculation from Eq.~\eqref{eq: numerical_model} assuming constant mode shape~\cite{stuart1958non}. At first glance, it looks like assuming constant mode shapes yields a closer prediction of the perturbation energy. However, a closer look shows that while calculation from Eq.~\eqref{eq: numerical_model} assuming varying mode shapes underestimates the steady state perturbation energy, the trend with varying freestream Mach number is captured correctly (while the trend is reversed if the mode shapes are assumed constant). The primary reason for underestimated steady state perturbation energies is a calculation of mean flow distortion using Eqs.~\eqref{eq: distorted-mean-flow}, which holds only at the steady state. Consequently, the mean flow distortion is overpredicted at each intermediate step of the model calculation. Physically, the time scale of mean-flow distortion is the same as the time scale which the perturbation energy increases, hence a more accurate model must integrate the governing Eqs.~\eqref{eq: distortion_gov} in time with the perturbation energy Eq.~\eqref{eq: numerical_model} along with the homogeneous Neumann boundary conditions far away from the boundary layer for the mean flow variables. 

\begin{table}[]
 \caption{Steady state energy in DNS, as calculated by using \citeauthor{stuart1958non}\cite{stuart1958non} (1958) and current model (Eq. \eqref{eq: numerical_model}). }
    \centering
    \begin{ruledtabular}
        \begin{tabularx}{\linewidth}{p{10mm}p{25mm}p{25mm}p{25mm}}
       
        Case & DNS  & Stuart\cite{stuart1958non} (1958)  & Eq. \eqref{eq: numerical_model} \\[1mm]
        \hline \\[.5mm]
         $\mathrm{M1}$ & $6.45\times 10^{-5}$ & $5.58 \times 10^{-5}$   & $9.17\times 10^{-6}$ \\ [1mm]
        
         M2 & $8.023\times 10^{-5}$  & $4.79 \times 10^{-5}$   & $1.00 \times 10^{-5}$  \\ [1mm]
         
         M3 & $9.716\times 10^{-5}$  & $3.87 \times 10^{-5}$  & $1.01\times 10^{-5}$ \\ [1mm]

        \end{tabularx}
    \end{ruledtabular}

    \label{tab: energy_compare_model}
\end{table}

\section{Conclusions}
\label{sec: conclusions}
We have studied the nonlinear saturation mechanism of the unstable Mack modes~\cite{mack1969boundary} in a hypersonic parallel flow boundary layer. The linearly unstable modes extract energy from the base flow via the Reynolds stress type production term. The production mechanism depends on the relative phase between the streamwise ($u$) and wall-normal velocity ($v$) fluctuations. This phase is significantly affected by the viscous effects close to the wall (wall layer) and near the critical line (critical layer) in the linear regime. Assuming $\mathrm{Re}_{\delta^{\star}} = \epsilon^{-2}$ for small $\epsilon$, the wall layer and the critical layer scales of $ \epsilon$ and $\epsilon^{2/3}$, respectively, yield viscous corrections in terms of the Airy's functions. Using these, we showed that the overall production of the perturbation energy is captured well as compared to the LST and DNS. Additionally, the perturbation energy production extends well beyond the relative sonic line (turning point). 

As the unstable mode keeps getting amplified, the nonlinear terms become significant due to increased amplitudes of the perturbations. These nonlinear terms result in two primary effects, the cascade of energy in higher harmonics (wave steepening) and the distortion of the mean flow. The energy in higher harmonics is less than $1\%$ of the total perturbation energy at the steady state. Additionally, the higher harmonics are generated coherently, that is, the phase speed remains constant. The mean flow distortion plays an important role in the nonlinear saturation. As the perturbations gain energy, the overall flow is decelerated, as a result of which the instability growth rates are reduced. This is accompanied by a reduction in the production term. After sufficient reduction, the production term matches the dissipation, and the perturbation energy stabilizes. The reduction in the production term is due to the change in relative phasing between the perturbations in $u$ and $v$ caused by the mean flow distortion. This highlights that the distortion in relative phasing must be modeled along with the mean flow distortion, unlike previous parallel flow studies~\cite{stuart1958non, stuart1960non, watson1960non, stewartson1971non} in which the reduction in production is attributed only to the mean gradient reduction, assuming the relative phasing to be constant ($\beta_3$ in Eq. 2.15 of \citeauthor{stuart1958non}~\cite{stuart1958non} cannot be treated as a constant and must be modeled as the mean flow distorts).

The present analysis yields a framework to study the high-speed parallel compressible shear flows transitioning to turbulence via hydrodynamic instabilities. \Revision{}{In 3D flows, the transition to turbulence is preceded by secondary instability, an oblique mode instability. The secondary instability can grow on the distorted mean flow saturated by primary instability (laminar +  mean flow distortion + saturated primary instability) either by fundamental resonance, \cite{zhu2020nonlinear, husmeier2007numerical} or subharmonic resonance \cite{adams1996subharmonic}.}  In a 3D saturation model, predicting correct distorted mean flow is extremely important for predicting the correct onset of secondary instabilities and breakdown to turbulence. \Revision{}{In 3D temporal DNS, transition follows the subharmonic resonance route. Additionally, the saturation of the base state will also be due to planar primary instability, thus allowing us to use the current model to calculate the correct distorted base state. The primary difference in 3D compared to 2D is the transition to turbulence through secondary instability instead of reaching a laminar saturated state, as seen in the 2D temporal case. In the spatial DNS, the frequency of the most unstable second mode will increase as we move downstream due to the growth of boundary layer \cite{zhu2022instability}. To model mean flow saturation in actual flow with broadband perturbations, we must account for a band of unstable frequencies to get the correct saturated mean flow. Otherwise, we can study the effect of one unstable frequency on the mean flow distortion through a controlled transition where we excite only one frequency in the boundary layer}.

\begin{acknowledgments}
We acknowledge the financial support received from Science and Engineering Research Board (SERB), Government of India under Grant No. SRG/2022/000728. We also thank IIT Delhi HPC facility for computational resources.
\end{acknowledgments}
\appendix
\section{Steady state in DNS}
\label{appA: steady_state_DNS}
\begin{figure}[!b]
    \centering
    \includegraphics[width=\linewidth]{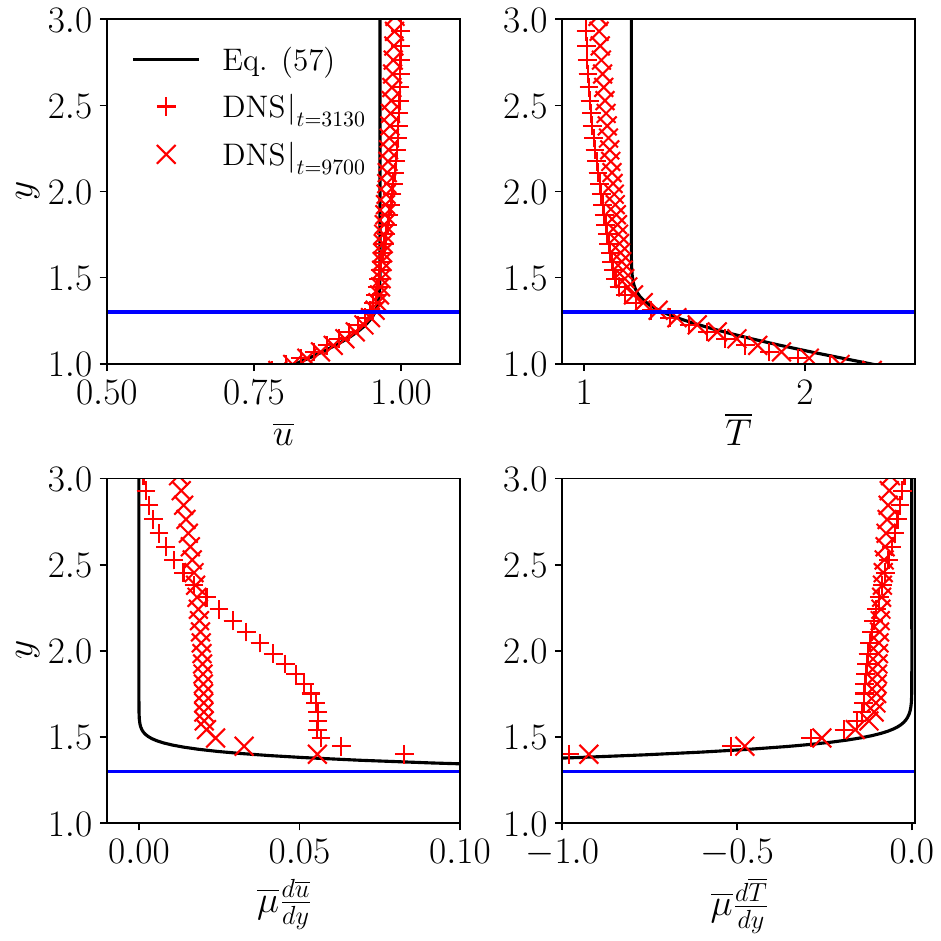}
    \put(-240,250){(a)}
    \put(-120,250){(b)}
    \put(-240,130){(c)}
    \put(-120,130){(d)}
    \caption{Comparison of the distorted mean as obtained from Eq.~\eqref{eq: numerical_model} at steady state and DNS at two later times (a) mean streamwise velocity (b) mean temperature (c) mean shear stress and (d) mean heat flux) for M1 case. The plots are focused near the edge of the boundary layer. Due to mean flow distortion, shear and heat flux generated above the boundary layer diffuse upwards $y>0$. Since this diffusion is small, the distorted mean flow from DNS takes a very long time to reach a true steady state. }
    \label{fig: distorted-mean-flow}
\end{figure}
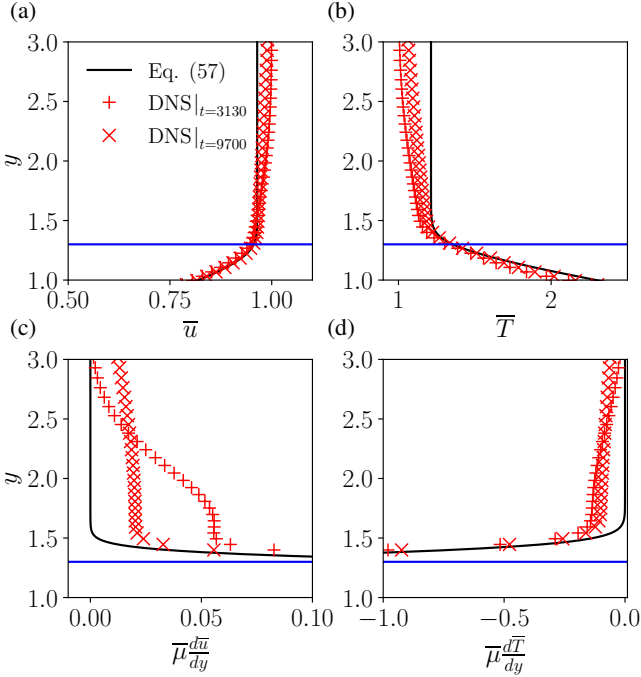
The distorted steady mean flow obtained from Eq.~\eqref{eq: numerical_model} and the DNS match well within the boundary layer (see Fig.~\ref{fig: steady_state_model}). However, we see a deviation outside the boundary layer. Outside the boundary layer, $f$ and $g$ quickly reach the freestream values. Moreover, the perturbation modes also decay exponentially. Hence, the unsteady mean flow distortion equations can be approximated as, 
\begin{subequations}\label{eq: unsteady-distorted-gov-eq}
    \begin{align}
    \frac{\partial \overline{\rho }\overline{u}}{\partial t} \approx& \frac{1}{\mathrm{Re}_{\delta^{\star}}}\frac{d}{dy}\left(\overline{\mu}\frac{d\overline{u}}{dy} \right),\\
    \frac{\partial \overline{\rho}\overline{T}}{\partial t}  \approx& \frac{\gamma (\gamma-1)M^2_{\infty}}{\mathrm{Re}_{\delta^{\star}}} \overline{\mu}\left(\frac{d \overline{u}}{dy}\right)^2 +\frac{\gamma}{\mathrm{Re}_{\delta^{\star}} \mathrm{Pr}} \frac{d}{dy}\left(\overline{\mu}\frac{d\overline{T}}{dy}\right).
 \end{align} 
\end{subequations}
In figure \ref{fig: distorted-mean-flow} (a) and \ref{fig: distorted-mean-flow} (b), we show the variation of  $\left(\overline{\mu}\frac{d\overline{u}}{dy}\right)$  and $\left(\overline{\mu}\frac{d\overline{T}}{dy}\right)$ along $y$ outside the boundary layer region. The quantity is non-zero and slowly moving towards zero with time, which suggests that the mean flow is still evolving with time but slowly. The distortion in shear and heat flux (mean wall-normal gradients of $\overline{u}$ and $\overline{T}$) outside the boundary layer due to the high amplitude modes diffuses till $y\to\infty$. \RevisionTwo{This diffusion is slow}. To save computational cost, we terminated our simulations as we saw the perturbation energy approaching a steady state value. However, to reach the true steady state, the simulations would have to be run for extremely long times. Since this diffusion of shear and heat flux takes place outside the boundary layer, the distortion in modes due to this diffusion is expected to be minimal. 
\begin{figure}
    \centering
    \includegraphics[width=\linewidth]{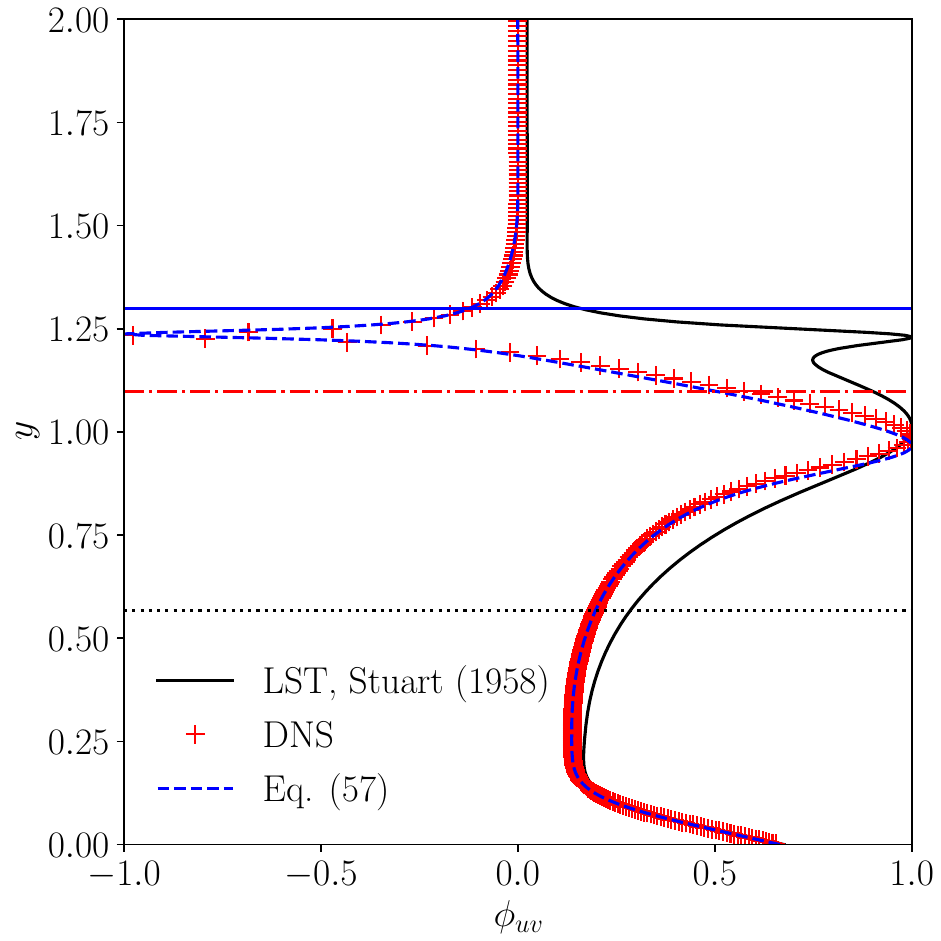}
    \caption{Phase difference between $u$ and $v$ ($\phi_{uv}$) defined in Eq.~\eqref{eq: phase_def} as obtained from Eq.~\eqref{eq: numerical_model} at steady state, DNS (at $t=9700$) perturbations in linear and non-linear saturated steady state for M1 case.}
    \label{fig: phase-compare}
\end{figure}
In Fig.~\ref{fig: phase-compare}, we show the phase $\phi_{uv}$ between $u$ and $v$ perturbations as evaluated from the DNS at steady state, the model Eq.~\eqref{eq: numerical_model} at steady state, the linear stability modes evaluates on $f$ and $g$. Clearly, in Stuart's~\cite{stuart1958non} model, the phase of the perturbations remains constant since the mode shapes are assumed constant. However, as discussed in Sec.~\ref{sec: mean-flow-distortion}, changes in mode shape are essential to capture the saturation and obtain the correct distorted state. Furthermore, Fig.~\ref{fig: phase-compare} also confirms that the DNS has reached a steady state since, with the given phasing, the growth rate of the perturbations (as obtained from the linear stability analysis on the distorted state) is very small ($\mathcal{O}\left(10^{-6}\right)$).

\section{Grid convergence}\label{appendix: convergence}

\begin{table}[t]
 \caption{Grid convergence for $M_{\infty}=5.5$. A uniform $\Delta y$ is used up to $y\le 1.4 $, and then it is exponentially coarsened till the sponge layer. }
 \begin{ruledtabular}
    \centering
    \begin{tabularx}{\linewidth}{p{15mm} p{10mm}p{10mm}p{15mm}p{15mm}p{25mm}}
         Case & $M_{\infty}$  & $\alpha$ & $N_x\times N_y$  & \Revision{$\Delta x$ }{$\Delta x^+$ } &\Revision{$\Delta y$ }{$\Delta y^+$ }
          \\[1mm]
        \hline \\[.5mm]
        $\mathrm{M3G1}$  & 5.5  & 2.14  & $60\times 75$ & \Revision{0.0979}{1.6053}  & \Revision{0.0269}{0.4416}   \\ [1mm]
        $\mathrm{M3G2}$  & 5.5  & 2.14  & $60\times 150$ &  \Revision{0.0979}{1.6053} & \Revision{0.01346}{0.2208}    \\ [1mm]
        $\mathrm{M3G3}$  & 5.5  & 2.14  & $120\times 75$ &  \Revision{0.0489}{0.8026} & \Revision{0.0269}{0.4416}   \\ [1mm]
        $\mathrm{M3G4}$  & 5.5  & 2.14  & $120\times 150$ &  \Revision{0.0489}{0.8026} &\Revision{0.01346}{0.2208}   \\ [1mm]
        $\mathrm{M3G5}$  & 5.5  & 2.14  & $180\times 75$ & \Revision{0.0326}{0.5351} & \Revision{0.0269}{0.4416}  \\ [1mm]
        $\mathrm{M3G6}$  & 5.5  & 2.14  & $180\times 150$ & \Revision{0.0326}{0.5351} & \Revision{0.01346}{0.2208}  \\ [1mm]
       
    \end{tabularx}
   \end{ruledtabular}
    \label{tab:grid_case}
\end{table}
\begin{figure}[!b]
    \centering
    \includegraphics[width=\linewidth]{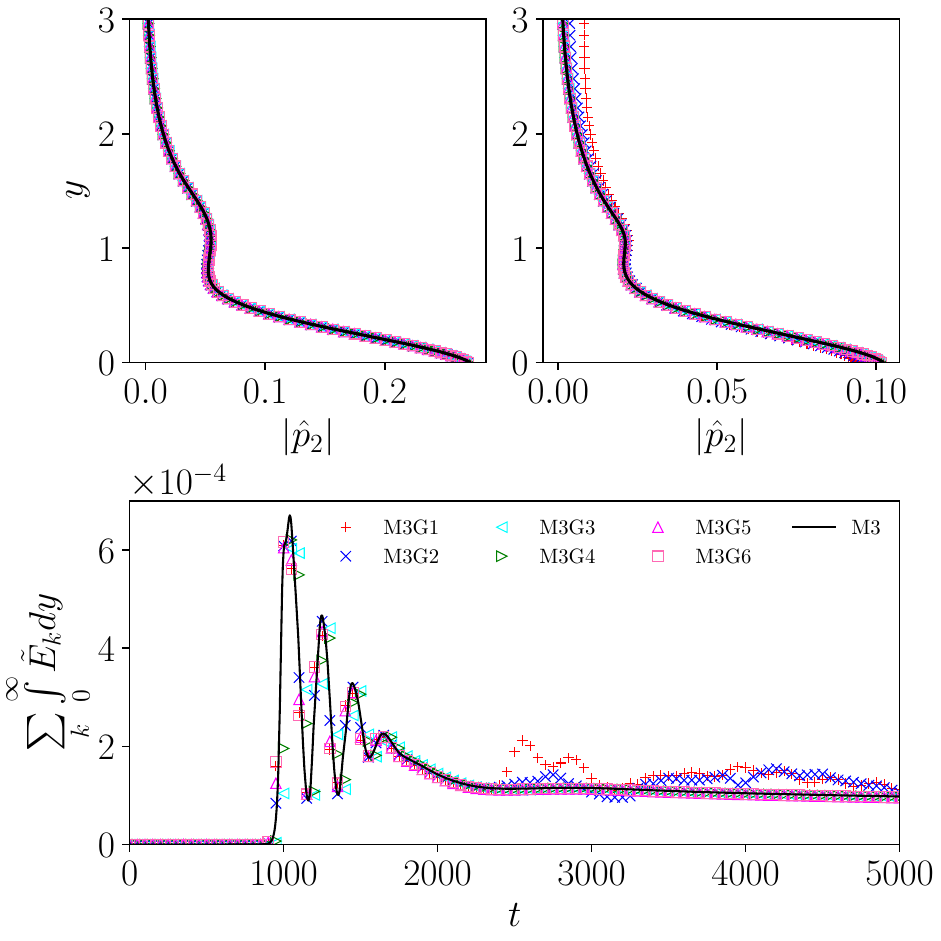}
    \put(-240, 250){(a)}
    \put(-120, 250){(b)}
    \put(-240, 125){(c)}
    \caption{Comparison of pressure profiles at a) peak amplitude and b) steady state is shown for different grids considered (for the legends, see Table \ref{tab:grid_case} and Table \ref{tab: parameters_table}). In c), we show the time series of the perturbation energy for all the grids. Except for M3G1 and M3G2, all cases overlap, highlighting grid convergence.} 
    \label{figPOF: grid convergence}
\end{figure}
To ascertain the grid independence of the results, we conducted a grid convergence test using six different grids (as shown in Table.~\ref{tab:grid_case}) for $M_{\infty}= 5.5$. Since $M_{\infty}= 5.5$ is the extreme case among all the cases considered (Table \ref{tab: parameters_table}) in terms of mean flow gradients and peak amplitude of perturbations, the grid independence of $M_{\infty}= 5.5$ will ensure the grid independence of others. Figure~\ref{figPOF: grid convergence} shows the pressure mode at peak and steady amplitude state for all the grids. We also show the time evolution of total perturbation energy. All grids considered follow the same evolution of perturbation energy until 2400 simulation time when cases M3G1 and M3G2 start deviating. The same convergence can be seen by plotting the pressure mode shape ($|\hat{p}_{k=2}|$) at two different time instants (peak and steady) for all the cases. Figure \ref{figPOF: grid convergence} (a) and \ref{figPOF: grid convergence} (b) confirm the grid convergence for a mesh with \Revision{$\Delta x\leq 0.0489$}{$\Delta x^+\leq 0.8026$} and \Revision{$\Delta y\leq 0.0489$}{$\Delta y^+\leq 0.4416$}. Ideally, the grid M3G3 is enough to simulate up to the steady state; we chose a finer grid in $y$ to extract more accurate higher-order derivatives of perturbation variables in $y$ at the post-processing stage. 

\Revision{}{
\section{Perturbation energy equation in nonlinear regime}\label{appendix: nonlinear_energy}
Using the decomposition in Eq.~\eqref{eq: decomposition-distortion}, we obtain the following weakly nonlinear equations,
\begin{align}
    &\frac{\partial \rho'}{\partial t} + \overline{\rho}\pfrac{u'}{x} + \overline{u}\pfrac{\rho'}{x} + \pfrac{\overline{\rho}v'}{y} + \epsilon_a\left(\pfrac{(\rho'u')}{x} + \pfrac{(\rho'v')}{y}\right) = 0\\
     &\pfrac{u'}{t} + \overline{u}\pfrac{u'}{x} + v'\frac{d\overline{u}}{dy} + \epsilon_a\left(u'\pfrac{u'}{x} + v'\pfrac{u'}{y}\right) +\frac{\left(1 - \epsilon_a\frac{\rho'}{\overline{\rho}}\right)}{\gamma M^2_\infty\overline{\rho}}\pfrac{p'}{x}\nonumber\\ &=D_u+ \epsilon_aD^{N}_u,\\
&\pfrac{v'}{t} + \overline{u}\pfrac{v'}{x}  + \epsilon_a\left(u'\pfrac{v'}{x} + v'\pfrac{v'}{y}\right) +\frac{\left(1 - \epsilon_a\frac{\rho'}{\overline{\rho}}\right)}{\gamma M^2_\infty\overline{\rho}}\pfrac{p'}{y}\nonumber\\ 
&
=D_v + \epsilon_aD^{N}_v\\
&\pfrac{p'}{t} + \overline{u}\pfrac{p'}{x} + \gamma \left(\pfrac{u'}{x} + \pfrac{v'}{y}\right) + \epsilon_a\left(u'\pfrac{p'}{x} + v'\pfrac{p'}{y}+\right.\nonumber \\
&\left. \gamma p'\left(\pfrac{u'}{x} + \pfrac{v'}{y}\right)\right) =  D_p + \epsilon_a D^N_p.
\end{align}
Substituting the spatial modal decomposition in Eq.~\eqref{eq: higher-harmonics-1}, multiplying with complex conjugates, and following the identical procedure as used for Eqs.~\eqref{eq: pertEnergyLinear}, we obtain Eq.~\eqref{eq: pertEnergyNonLinear} with the following triadic terms in the nonlinear energy exchange term $\mathcal{N}_k$, 
\begin{align}
&\mathcal{N}_k=-\epsilon_a\overline{\rho}\left(\Re\sum_{p+q=k}\left(i\tilde{u}^*_k\tilde{u}_pq\tilde{u}_q + \tilde{u}^*_k\tilde{v}_p\frac{\partial \tilde{u}_q}{\partial y} - \frac{iq\tilde{u}^*_k\tilde{\rho}_p\tilde{p}_q}{\gamma M^2_\infty\overline{\rho}^2}\right)\right) - \nonumber\\&
\epsilon_a\overline{\rho}\left(\Re\sum_{p+q=k}\left(\tilde{v}^*_k\tilde{u}_piq\tilde{v}_q + \tilde{v}^*_k\tilde{v}_p\frac{\partial \tilde{v}_q}{\partial y} - \frac{\tilde{v}^*_k\tilde{\rho}_p}{\gamma M^2_\infty\overline{\rho}^2}\pfrac{\tilde{p}_q}{y}\right)\right) - \nonumber\\
 &\frac{\epsilon_a\Re}{(\gamma M_\infty)^2}\left(\sum_{p+q=k}\left(\tilde{p}^*_k\tilde{u}_piq\tilde{p}_q + \tilde{p}^*_k\tilde{v}_p\pfrac{\tilde{p}_q}{y}\right)\right)-\nonumber\\
 &\frac{\epsilon_a\Re}{\gamma M^2_\infty}\left(\sum_{p+q=k}\left(\tilde{p}^*_k\tilde{p}_piq\tilde{u}_q + \tilde{p}^*_k\tilde{p}_p\pfrac{\tilde{v}_q}{y}\right)\right)+\nonumber\\&\epsilon_a\Re\left(\overline{\rho}\tilde{u}^*_kD^N_{\tilde{u}} + \overline{\rho}\tilde{v}^*_kD^N_{\tilde{v}} + \frac{\tilde{p}^*_kD^N_{\tilde{p}}}{(\gamma M_\infty)^2}\right).
\end{align}
The triadic terms have coefficients corresponding to the base flow multiplying terms corresponding to three perturbation quantities. Hence, at the leading order, the energy exchange for each $k^{th}$ mode is from the base flow and at the first higher order, the interaction is with the other modes. For brevity, we have not provided the detailed expressions of the nonlinear thermoviscous terms $D^N_u$, $D^N_v$, and $D^N_p$ which can be derived after considering $\mathcal{O}(\epsilon^2_a)$ terms in the diffusion terms.
}

\bibliography{aipsamp}

\end{document}